\documentclass[twocolumn, floatfix]{aastex61}

\usepackage[T1]{fontenc}
\usepackage{ae,aecompl,afterpage}

\newcommand\aastex{AAS\TeX}

\received{June 24, 2019}
\revised{\today}
\accepted{ZZZZ}

\shorttitle{\aastex\ TDGs in HCGs: HCG 26, 91, and 96}
\shortauthors{B. Nikiel-Wroczy\'nski}

\begin{document} 

\title{Searching for the magnetised Tidal Dwarf Galaxies in Hickson Compact Groups: HCG 26, 91, and 96}

\correspondingauthor{B\l a\.zej Nikiel-Wroczy\'nski}
\email{blazej.nikiel$\_$wroczynski@uj.edu.pl}

\author{B\l a\.zej Nikiel-Wroczy\'nski}
\affil{Astronomical Observatory of the Jagiellonian University, 
ul. Orla 171, 30-244 Krak\'ow, Poland}
\affil{Leiden Observatory, Leiden University,  
Oort Gebouw, PO Box 9513, NL-2300 RA Leiden, the Netherlands}

\begin{abstract}

In this work, archive 1.4 and 4.86\,GHz radio continuum data from the VLA were re-reduced and, together with the 1.4\,GHz maps from the NVSS, investigated for the presence of a detectable, non-thermal continuum radio emission that could be associated with the TDG candidates in HCG\,26, 91, and 96. Radio emission highly coincident with the optical and $\rm H_{\alpha}$ emission maxima of the TDG candidate HCG\,91i (estimated physical separation of less than 150\,pc) was revealed. Should this emission be intrinsic to this object, it would imply the presence of a magnetic field as strong as 11--16\,$\mu$G -- comparable to that found in the most radio-luminous, star-forming dwarf galaxies of non-tidal origin. However, the star formation rate derived for this object using the radio flux is about two orders of magnitude higher, than the one estimated from the $\rm H_{\alpha}$ data. Analysis of the auxiliary radio, ultraviolet and infrared data suggests that either the radio emission originates in a background object with an aged synchrotron spectrum (possibly a GHz-peaked source), or the $\rm SFR_{H_{\alpha}}$ estimate is lower due to the fact that it traces the most recent star formation, while most of the detected radio emission originated when what is known as HCG\,91i was still a part of its parent galaxy. The latter scenario is supported by a very large stellar mass derived from 3.6 and 4.5\,$\mu$m data, implying high star formation in the past.
   
\end{abstract}

\keywords{
galaxies:dwarfs -- galaxies:groups:individual:HCG26 -- galaxies:groups:individual:HCG91 -- galaxies:groups:individual:HCG96 -- galaxies:interactions -- galaxies:magnetic fields
          }

\section{Introduction}
\label{section:introduction}

The idea that small, self-gravitating entities can form out of the debris left by collisions and close passages of "normal" galaxies is a relatively new one, as it was proposed only about 60\,years ago \citep{zwicky56}. Zwicky analysed several aspects of the galactic interactions, e.g. the emergence of large tails, like that of the Leo Triplet, and on the basis of the disputed cases, came to a conclusion that defines the unique character of the tidal dwarf galaxies (TDG). However, albeit novel and (as turned out eventually) correct, Zwicky's idea did not gain much attention for more than 30 years. This was mainly due to the technical limitations of the that-day instruments that simply could not achieve sensitivity required to trace the weak light produced by the TDGs. In 1992, \citeauthor{mirabel92} have announced a detection of such an object at the tip of the tidal tail of a famous merging galaxy pair, the Antennae. With this first discovery, the era of studying the TDGs has began.\\

The following years have brought a number of case studies of the TDGs, with examples of such objects detected eg. in Arp\,245 \citep{brinks04}, in the Virgo Cluster \citep{duc07}, or in galaxy groups, like in M81 \citep{makarova02}, or in a number of Hickson Compact Groups (HCG, \citealt{hickson82}), by \citet{hunsberger96}. An important warning has also been formed: as only about 50\% of the TDGs can survive for as long as the Hubble time \citep{bournaud10}, and usually self-gravitation is only assumed, not proven, it is advisable to call these numerous detections as "candidates" (TDGc), being probable, but not certain, galaxies \citep{kaviraj12}. The growing database of supposed dwarfish galaxies with a tidal origin was calling for a statistical approach, in which typical parameters and traits of TDGs could be analysed. This was done by \citet{kaviraj12}, who used the SDSS DR6 data \citep{sdss6} and built a sample of more than 3000 nearby (not further than $z=0.1$) galaxy mergers that could possibly host TDG candidates. Inside these systems, as many as 405 candidate objects have been identified.\\

Studies on TDGs in galaxy groups constitute a separate and growing chapter in the history of their observations. The TDG formation efficiency drops down rapidly with the increasing density of the host systems \citep{kaviraj12}. They are much more scarce in galaxy groups than in pairs, and probably are eager to evolve differently. Complicated dynamics of the multi-galaxy systems increases the possibility that a newly born TDG candidate/progenitor will be absorbed in the subsequent galactic collisions (like it could have happened eg. in case of the Leo Triplet -- \citealt{wezgowiec12}). Despite this impediment, studies carried out on the HCGs show that these systems can be quite abundant with TDG candidates. In particular, HCG\,92 -- the famous Stephan's Quintet \citep{stephan77} -- has been concluded to host at least 20 candidate objects \citep{hunsberger96}, two of them (denoted SQ--A and SQ--B; \citealt{xu03}) being visible in multiple spectral regimes. A recent study by \citet{eigenthaler15} shows that other HCGs are also accompanied with TDG candidates, and some of these groups can host a significant number of them (eg. HCG\,91).\\

When discussing the TDG candidates found in galaxy groups, two of them should be especially mentioned: SQ--A, and SQ--B which are located in the tidal tail of NGC\,7319, a member galaxy of the Stephan's Quintet. What is special about these objects is that they emit in the radio continuum \citep{xu03}. This emission has a non-thermal character, and there are hints that it is partially polarised in case of the latter one \citep{bnw13B}, signifying the existence of a detectable magnetic field, probably at least ordered (if not genuinely regular), inside this objects. Albeit a possibility that the matter forming a TDG is magnetised is certainly not an exotic one -- most of the spiral galaxies host relatively strong magnetic fields \citep{niklas95}, and these are mostly spirals that serve as progenitors to TDG candidates (TDGc, \citealt{kaviraj12}) -- SQ--A and SQ--B remain the single known examples of the magnetised TDGc so far. Hints for the radio emission from a TDGc in Leo Triplet \citep{bnw13A} have turned out to be caused by a background source smeared by the large beam of the single-dish observations used in that study \citep{bnw14B}. Attempts to detect radio emission in other TDGc -- eg. in the Antennae, K.~Chy\.zy, \citetext{priv. comm.} -- have turned out to be fruitless so far. \\

In general, detection of the magnetic field inside, or around a TDGc should not be very surprising. Conditions favourable for their formation, like the presence of spiral galaxies, or effective processes of star forming, are preferable for the magnetic fields to be amplified, too. However, magnetic fields found in dwarfish galaxies are usually weaker, than those found in the spiral ones: whereas the median strength of the magnetic field in spiral galaxies is $9 \pm 1.3 \mu$G \citep{niklas95}, values as low as a few microgauss are usually reported for the dwarf ones (see \citealt{chyzy11} and references therein). Strong magnetic fields are found only in the so-called starburst dwarf galaxies (eg. NGC\,1569, \citealt{kepley10}, or NGC\,4449, \citealt{chyzy00}), and can be considered an exception to this rule. Alas, the sample of the studied TDG candidates is still too scarce to investigate the role and general properties of the magnetic fields inside them, and it is desirable to enlarge it.\\

In this paper, the results of investigating the radio maps of those HCGs that have been marked by \citet{eigenthaler15} as possible hosts of TDG candidates are presented. Archive radio data from the Very Large Array (VLA), and the NRAO VLA Sky Survey (NVSS; \citealt{nvss}) were analysed for each of these systems, revealing radio emission possibly associated with HCG\,91i both at 1.4 and 4.86\,GHz. The paper is organised as follows: Sec.~\ref{section:data} contains the basic information on the data used, as well as information about their processing, Sec.~\ref{section:results} describes the radio maps of the studied groups and relates them to the list of TDG candidates compiled by \citet{eigenthaler15}. Sec.~\ref{section:discussion} discusses the arguments in favour and against possible connection between the detected radio emission and HCG\,91i, as well as evaluates different explanations for the detected signal, while Sec.~\ref{section:conclusions} summarises the findings of this article.

\section{Observations and Data Reduction}
\label{section:data}

Three systems that were assumed to be hosts of possible TDG candidates by \citet{eigenthaler15} were taken into account. TDGs are (physically) small objects, so their expected angular size is also rather small; no extended radio emission from them is likely to be revealed (unless radio data of subarcsecond resolution are used). As many of them form close to their host galaxies \citep{kaviraj12}, they are prone to be mistaken with other similar, but not self-gravitating entities, like e.g. $\rm H_{II}$ regions. Therefore, in order to prevent beam smearing of the emission from several entities into one, only loose-configuration, archive VLA data (at most B-configuration at 1.4\,GHz, and C-configuration at 4.86\,GHz) were chosen. Table \ref{tab:data} lists the basic information on the selected datasets.\\

All of the data used were imported into the Astronomical Image Processing System (\textsc{aips}) and calibrated following the standard continuum UV data calibration procedure. Final images were deconvolved using the \textsc{clean} algorithm implemented in the task \textsc{imagr}, and corrected for the primary beam attenuation. Later on, the geometry of the maps of the Total Power (TP) radio emission has been transformed using Everett interpolation to make it consistent with that of the cutouts from an optical sky survey using the task \textsc{hgeom}, flux measurements were taken, and final images were created. In addition, NVSS data were also used to confirm if the possible detections -- which were expected to be weak, thus possibly ambiguous -- are represented also in the more sensitive, low-resolution map.

%====================================
\begin{table*}[htp]
\caption{\label{tab:data}Basic information on the interferometric datasets used in this study.}
\begin{center}
\begin{tabular}{l|ccccc|ccccc}
\hline
\hline
HCG	&	L-band	&			&						&							&			&	C-band	&			&						&							&			\\
No.	&	Project	&	TOS [s]	&	Resolution			&	Noise [$\mu$Jy/beam]	&	Conf.	&	Project	&	TOS	[s]	&	Resolution			&	Noise [$\mu$Jy/beam]	&	Conf.	\\	
\hline
26	&	AM344	&	930		&	$8.9'' \times 5.8''$	&	70						&	A/B		&	AM219	&	340		&	$1.8'' \times 1.6'' $	&	30						&	C		\\
91	&	AT149C	&	440		&	$12.8''\times 8.1''$	&	250						&	B/C		&	AC345B	&	190		&	$8.3'' \times 3.4''	$&	125						&	C		\\
96	&	AS267	&	1350	&	$5.4'' \times 4.3''$    &	120						&	B		&	AM219	&	350		&	$1.2'' \times 0.7'' $	&	75						&	A		\\
\hline
\end{tabular}
\end{center}
\end{table*}
%====================================

\section{Results}
\label{section:results}

\begin{figure}[htp]
\centering
	\includegraphics[width=0.38\textwidth]{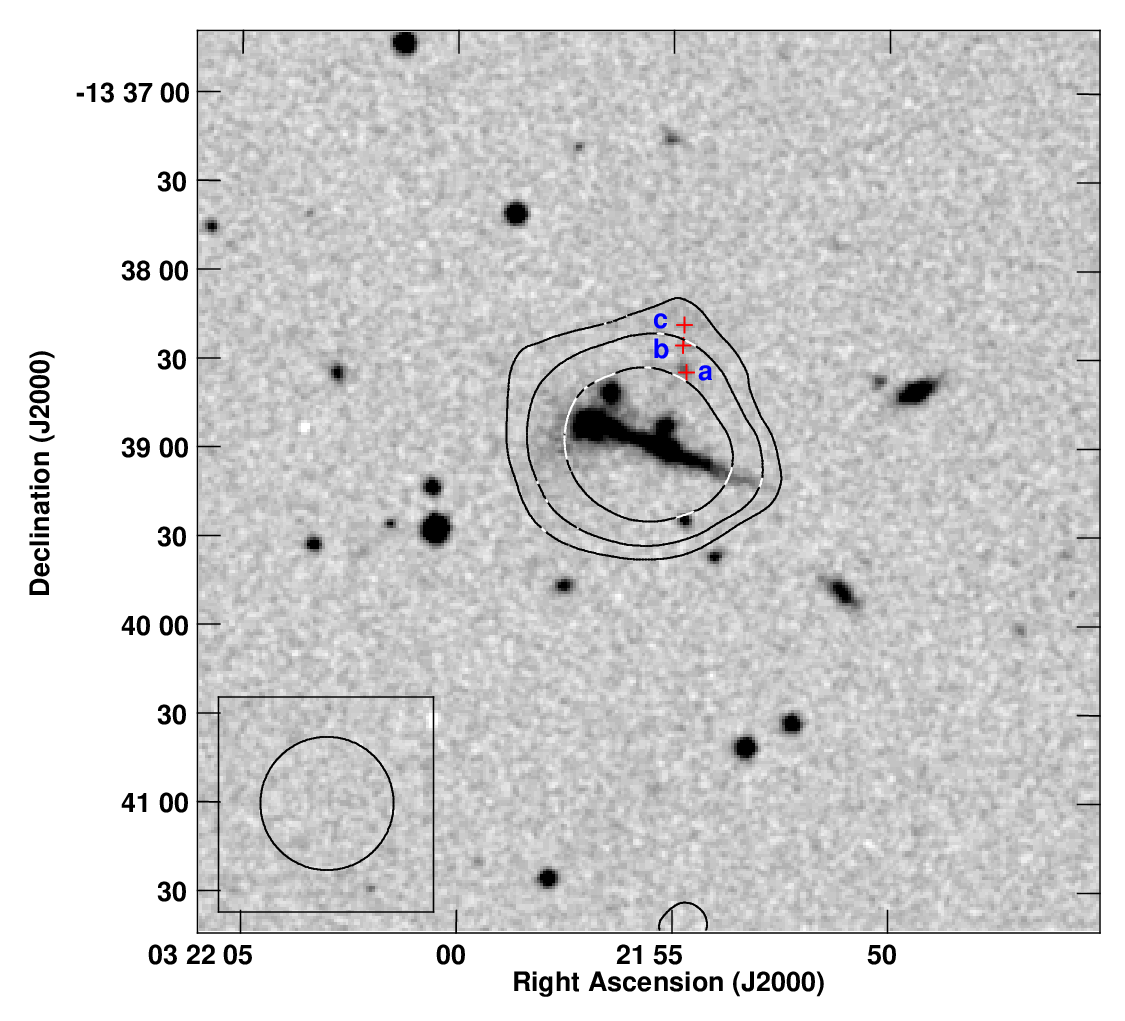}
    \includegraphics[width=0.38\textwidth]{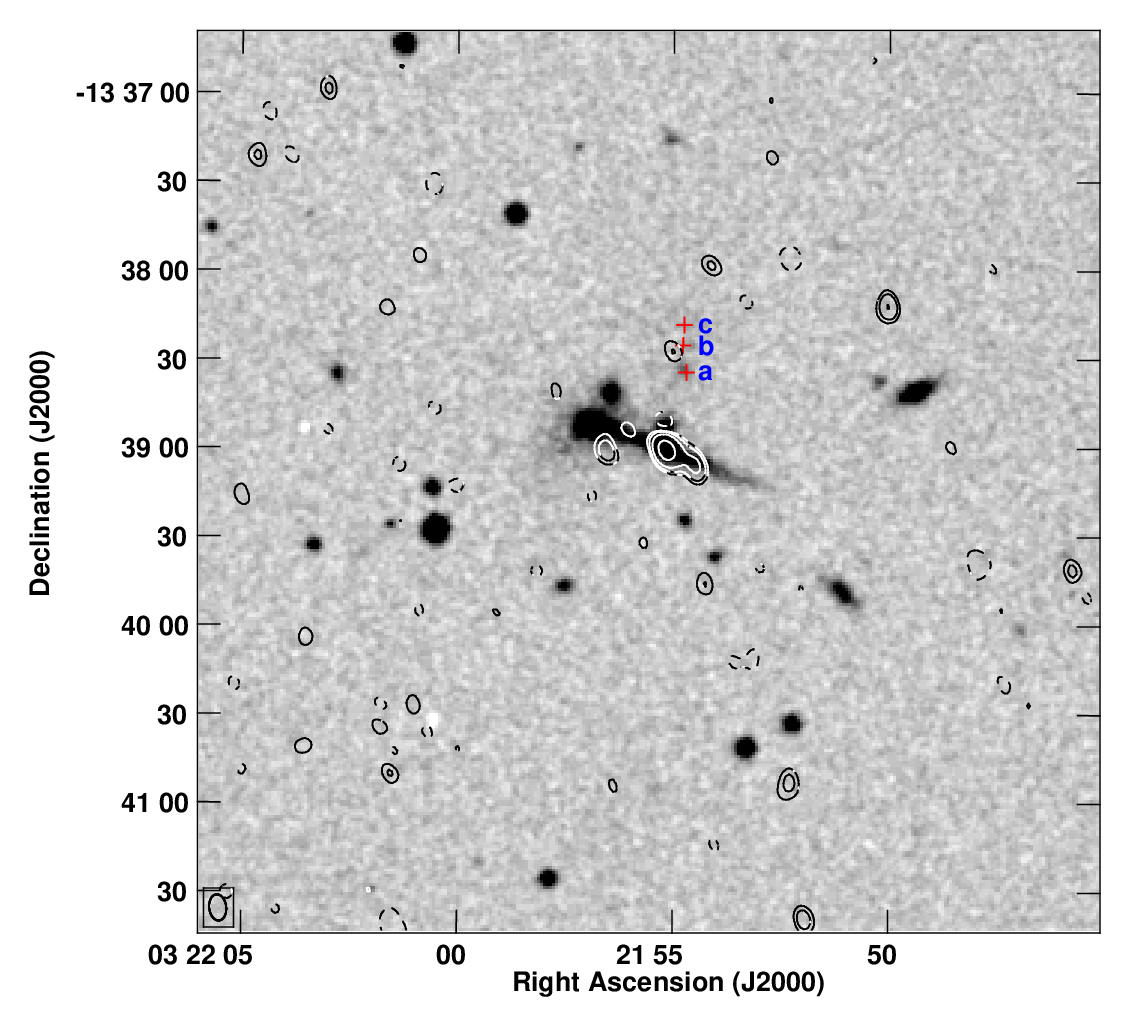}
    \includegraphics[width=0.38\textwidth]{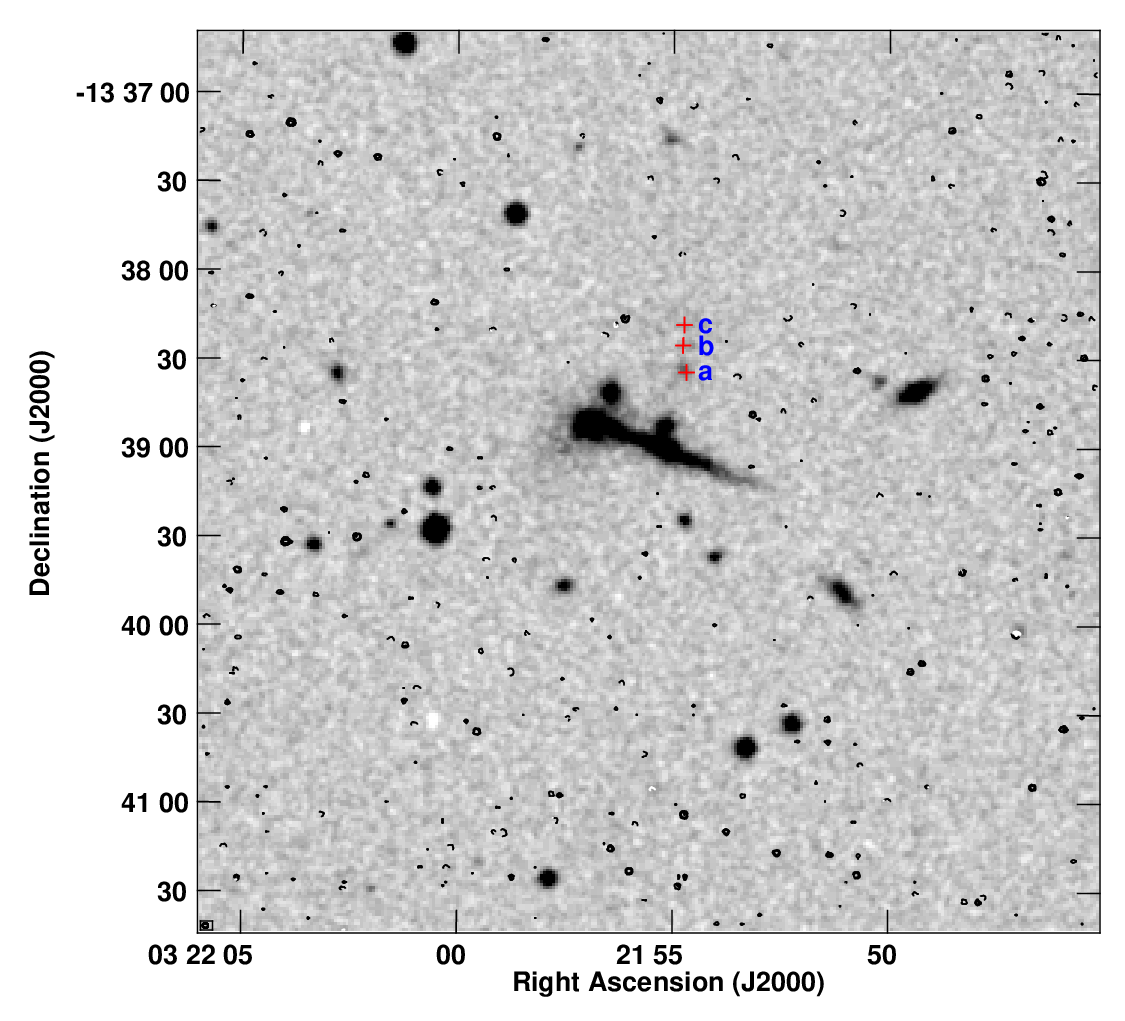}
\caption{\label{fig:h26_main}Maps of the radio emission of HCG\,26. Radio contours overlaid on a POSS-II Johnson-R filter map. The levels are 3, 5, 10, 25, 50 $\times$ r.m.s. noise level. \textbf{Upper panel:} NVSS map; r.m.s. noise level of 450\,$\mu$Jy/beam, angular resolution of 45\arcsec$\times$45\arcsec. \textbf{Middle panel:}  1.4\,GHz emission; r.m.s. noise level of 70\,$\mu$Jy/beam, angular resolution of 8.9\arcsec$\times$5.8\arcsec. \textbf{Lower panel:} 4.86\,GHz emission; r.m.s. noise level of 30\,$\mu$Jy/beam, angular resolution of 1.8\arcsec$\times$1.6\arcsec.}     
\end{figure}

\begin{figure}[htp]
\centering
	\includegraphics[width=0.38\textwidth]{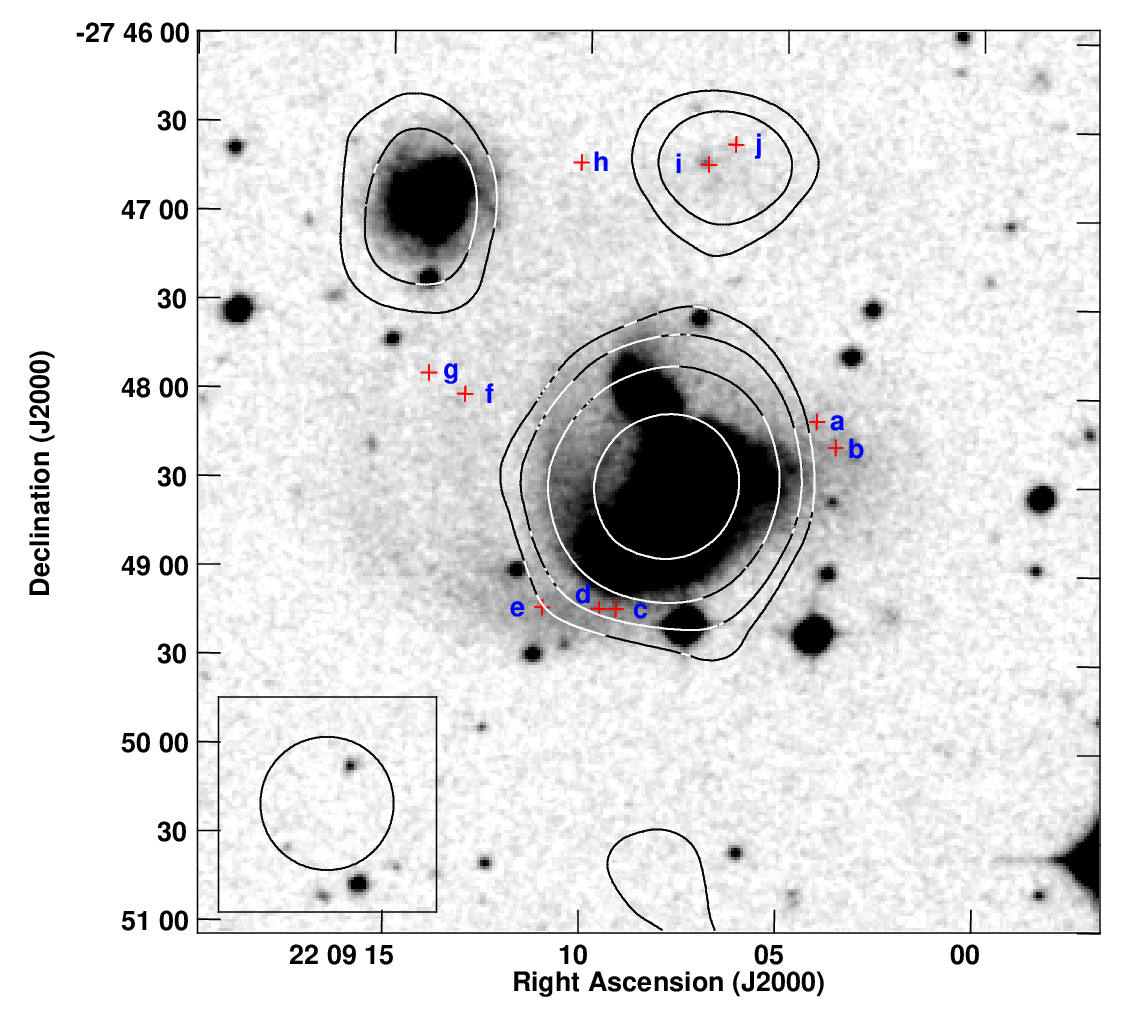}
    \includegraphics[width=0.38\textwidth]{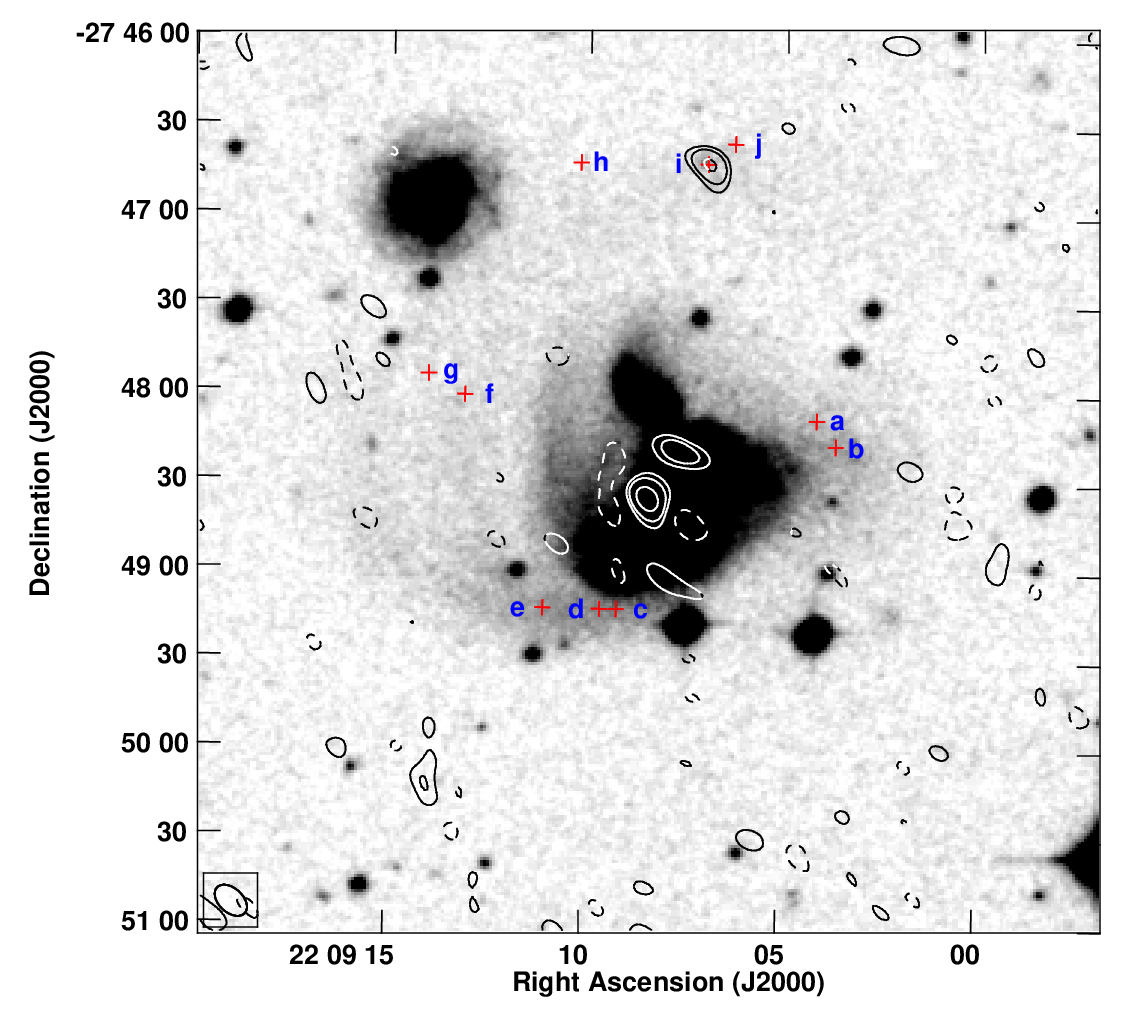}
    \includegraphics[width=0.38\textwidth]{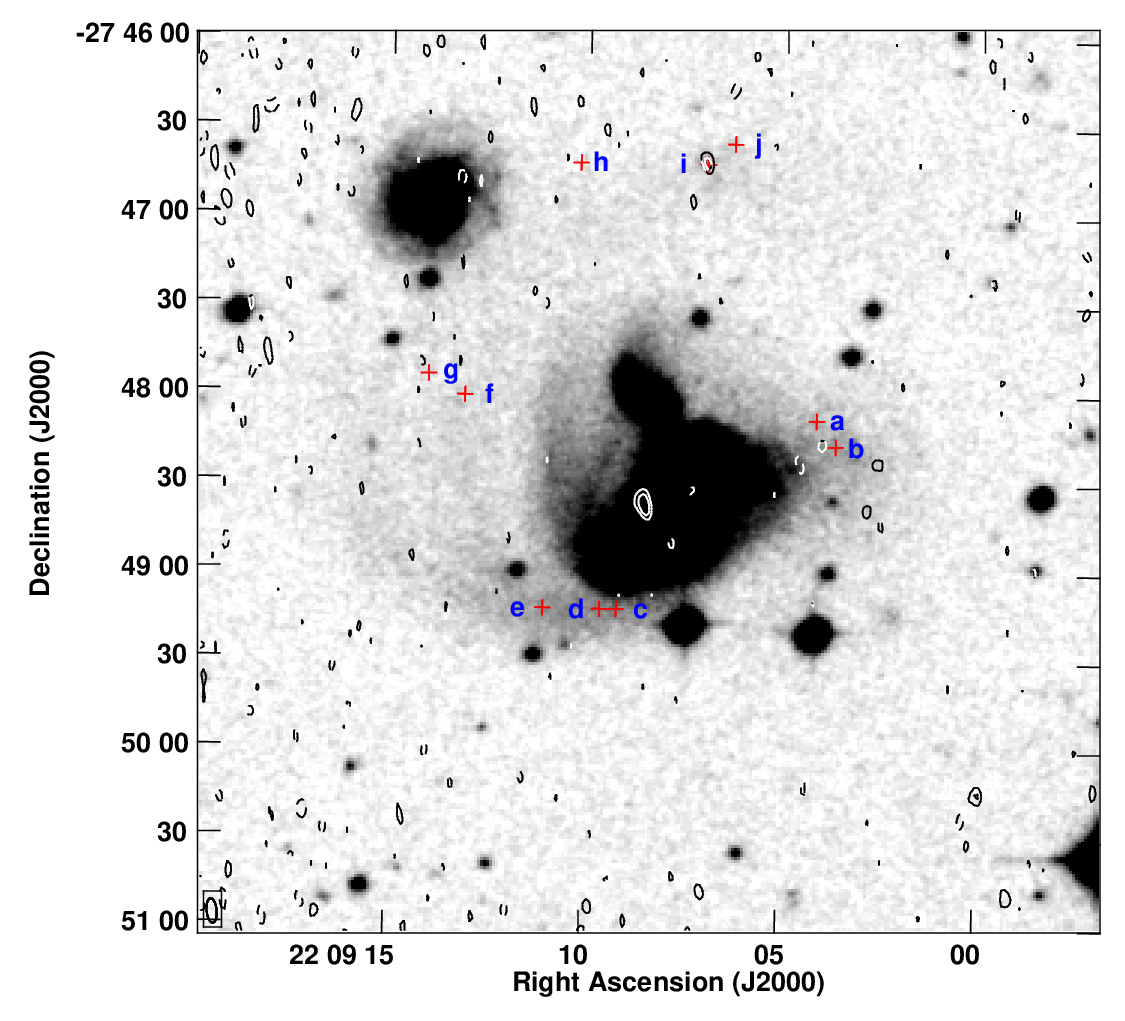}
\caption{\label{fig:h91_main}Maps of the radio emission of HCG\,91. Radio contours overlaid on a POSS-II Johnson-R filter map. The levels are 3, 5, 10, 25, 50 $\times$ r.m.s. noise level. \textbf{Upper panel:} NVSS map;  r.m.s. noise level of 450\,$\mu$Jy/beam, angular resolution of 45\arcsec$\times$45\arcsec. \textbf{Middle panel:}  1.4\,GHz emission; r.m.s. noise level of 250\,$\mu$Jy/beam, angular resolution of 12.8\arcsec$\times$8.1\arcsec. \textbf{Lower panel:} 4.86\,GHz emission; r.m.s. noise level of 125\,$\mu$Jy/beam, angular resolution of 8.3\arcsec$\times$3.4\arcsec.}        
\end{figure}

\begin{figure}[htp]
\centering
	\includegraphics[width=0.4\textwidth]{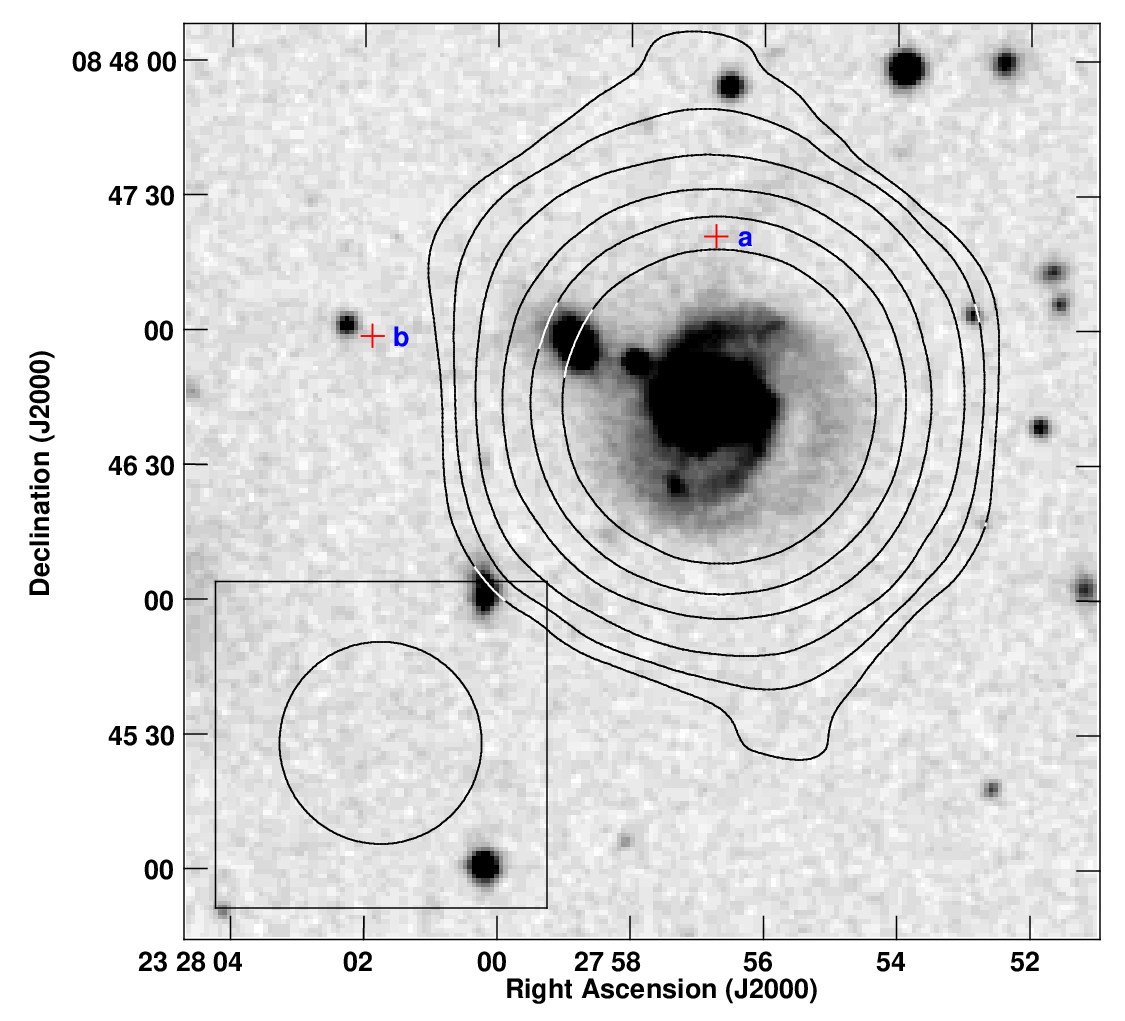}
    \includegraphics[width=0.4\textwidth]{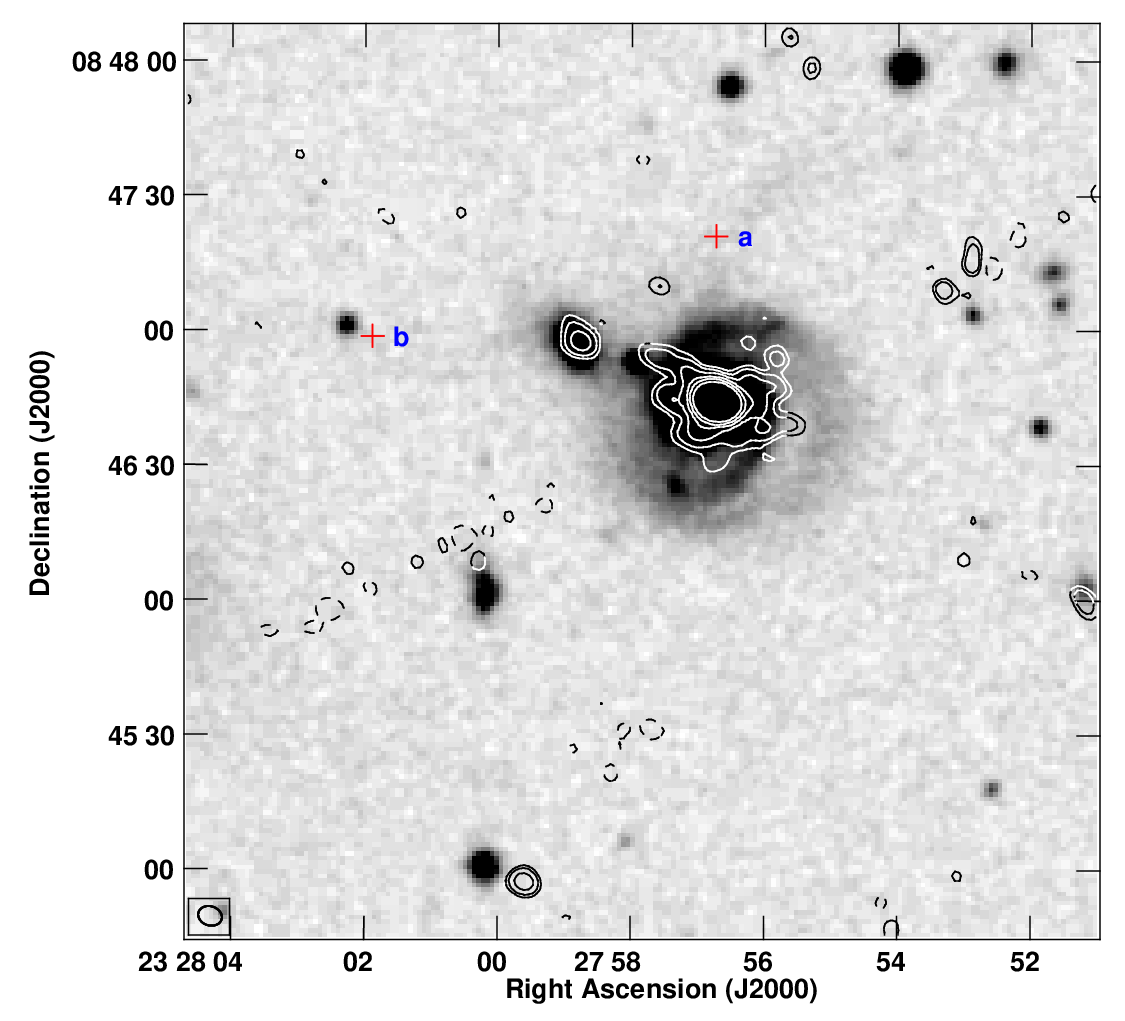}
    \includegraphics[width=0.4\textwidth]{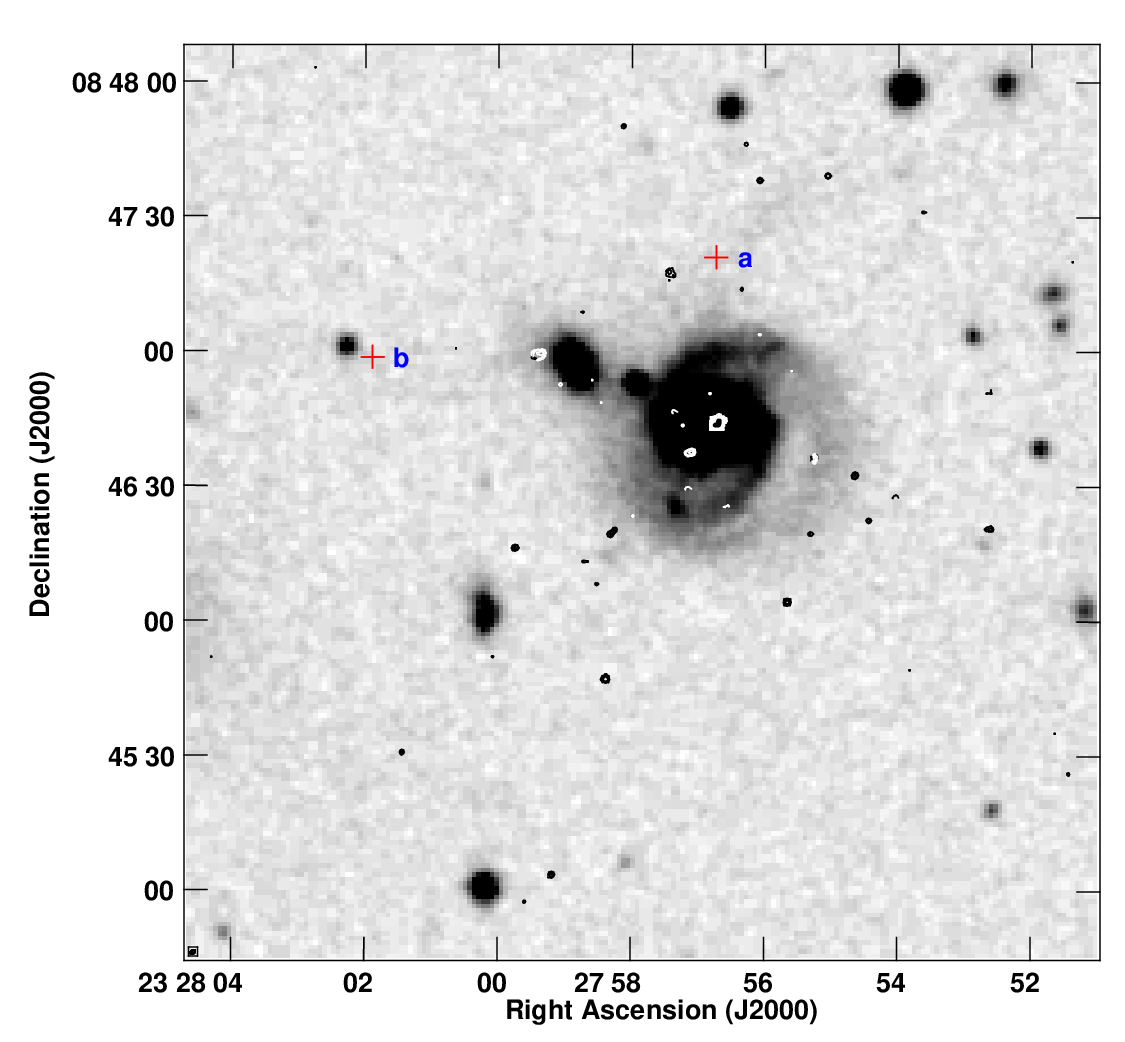}
\caption{\label{fig:h96_main}Maps of the radio emission of HCG\,96. Radio contours overlaid on a POSS-II Johnson's red map. The levels are 3, 5, 10, 25, 50 $\times$ r.m.s. noise level. \textbf{Upper panel:} NVSS map; r.m.s. noise level of 450\,$\mu$Jy, angular resolution of 45\arcsec$\times$45\arcsec. \textbf{Middle:}  1.4\,GHz emission; r.m.s. noise level of 120\,$\mu$Jy, angular resolution of 5.4\arcsec$\times$4.3\arcsec. \textbf{Bottom:} 4.86\,GHz emission; r.m.s. noise level of 90\,$\mu$Jy, angular resolution of 1.2\arcsec$\times$0.7\arcsec.}     
\end{figure}

For each of the three systems studied, maps of the radio emission at 1.4, and 4.86\,GHz were made. They are included in Fig.~\ref{fig:h26_main} (HCG\,26), Fig.~\ref{fig:h91_main} (HCG\,91), and Fig.~\ref{fig:h96_main} (HCG\,96). Each of them is a three-image panel, with the upper part presenting the NVSS data, the middle -- 1.4\,GHz high-resolution archive data, and the bottom one -- high-resolution archive data taken at 4.86\,GHz. Area covered by the each of the radio maps was chosen in such a way that all of the TDG candidates are included. Throughout the paper, spatial distances were calculated according to \citet{hickson92}.\\

%=============================
%=============================

\begin{figure}[htp]
\centering
    \includegraphics[width=0.4\textwidth]{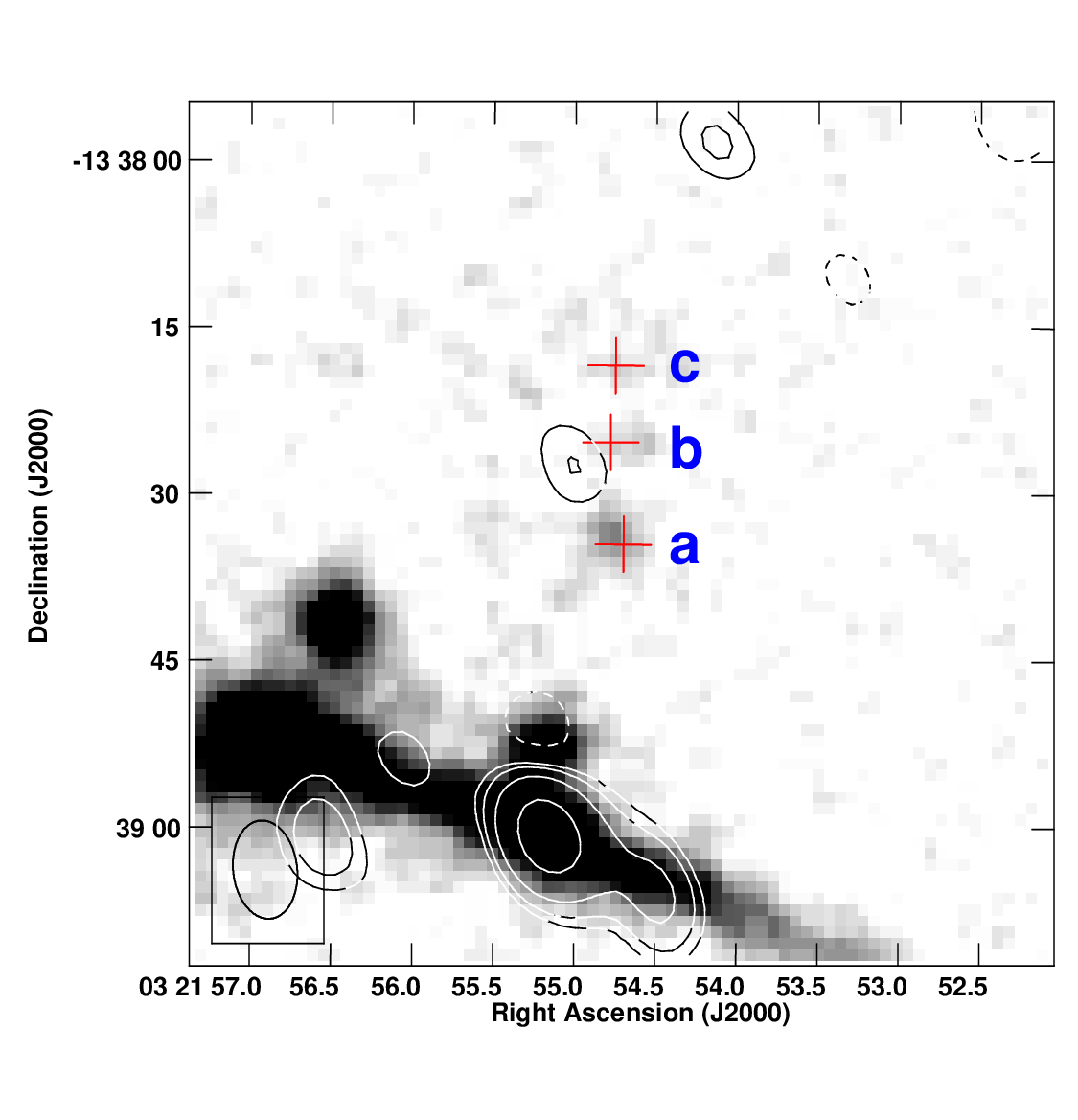}
\caption{\label{fig:h26_int}Map of the radio emission of HCG\,26 at 1.4\,GHz in the vicinity of TDG candidates \textbf{a}, \textbf{b}, and \textbf{c}. Radio contours overlaid on a (high-contrast) cutout of the POSS-II Johnson-R filter map. The levels are 3, 5, 10, 25 $\times$ r.m.s. noise level of 70\,$\mu$Jy. The angular resolution is 8.9\arcsec$\times$5.8\arcsec.}     
\end{figure}

First of the objects, HCG\,26, is well visible in the NVSS, with an extension of the radio contour that encompasses all three TDG candidates (Fig.~\ref{fig:h26_main}, upper panel). The 1.4\,GHz archive data (Fig.~\ref{fig:h26_main}, middle panel) show that most of this emission is likely to originate within the central galaxy, with only an isolated patch -- barely exceeding the 3\,r.m.s. level -- located close to the TDG candidates. However, a close inspection (Fig.~\ref{fig:h26_int}) reveals that the maximum of the radio emission at 1.4\,GHz ($\rm 0.34 \pm 0.07 $ mJy) is displaced from the optical structure; the distance between the aforementioned patch and the closest candidate (HCG\,26b) is comparable to the size of the telescope beam. At 4.86\,GHz (Fig.~\ref{fig:h26_main}, lower panel), none of the aforementioned structures is still visible. Therefore, it is concluded that the detected emission is not associated with any of the TDG candidates.\\

%=============================
%=============================

\begin{figure}[htp]
\centering
    \includegraphics[width=0.38\textwidth]{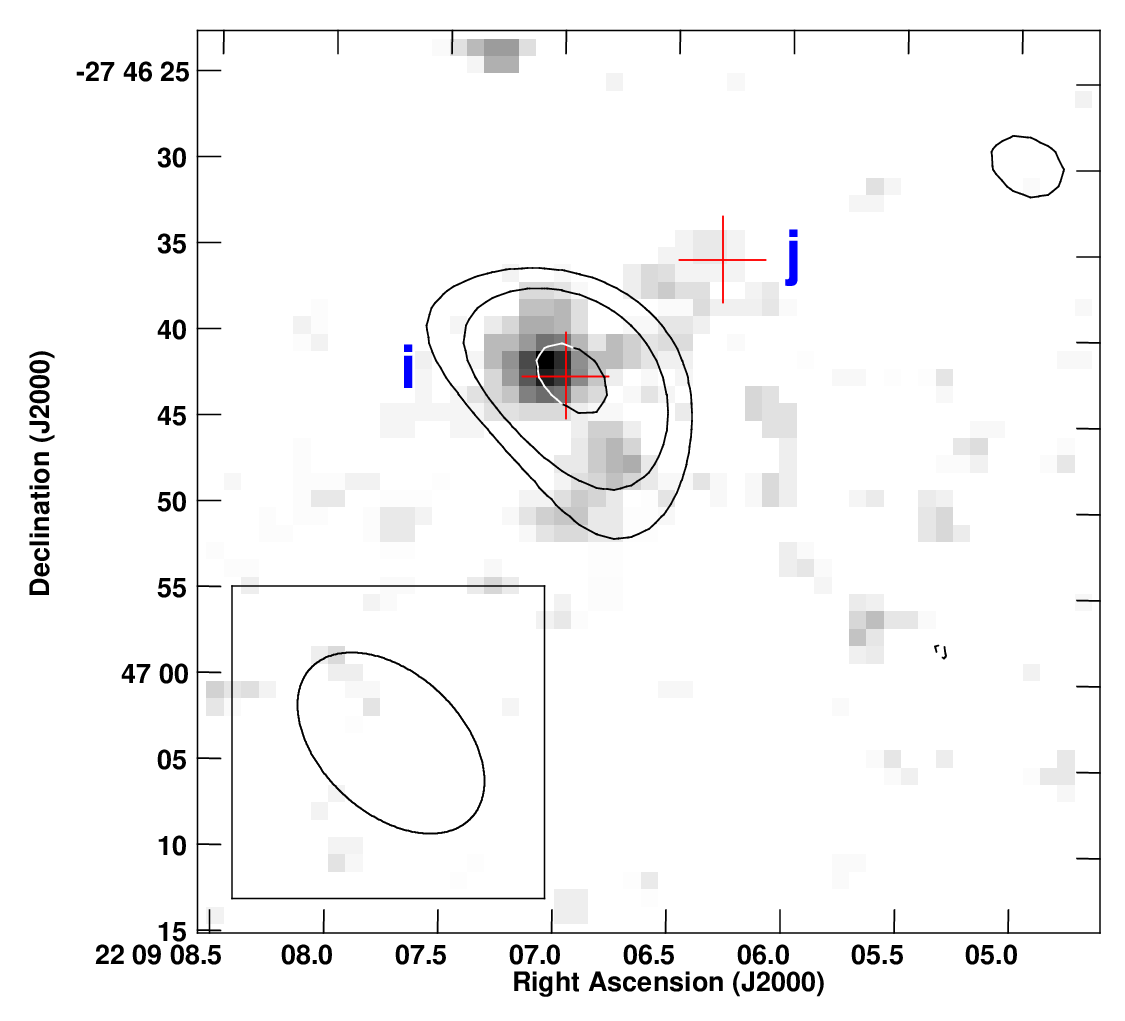}
    \includegraphics[width=0.38\textwidth]{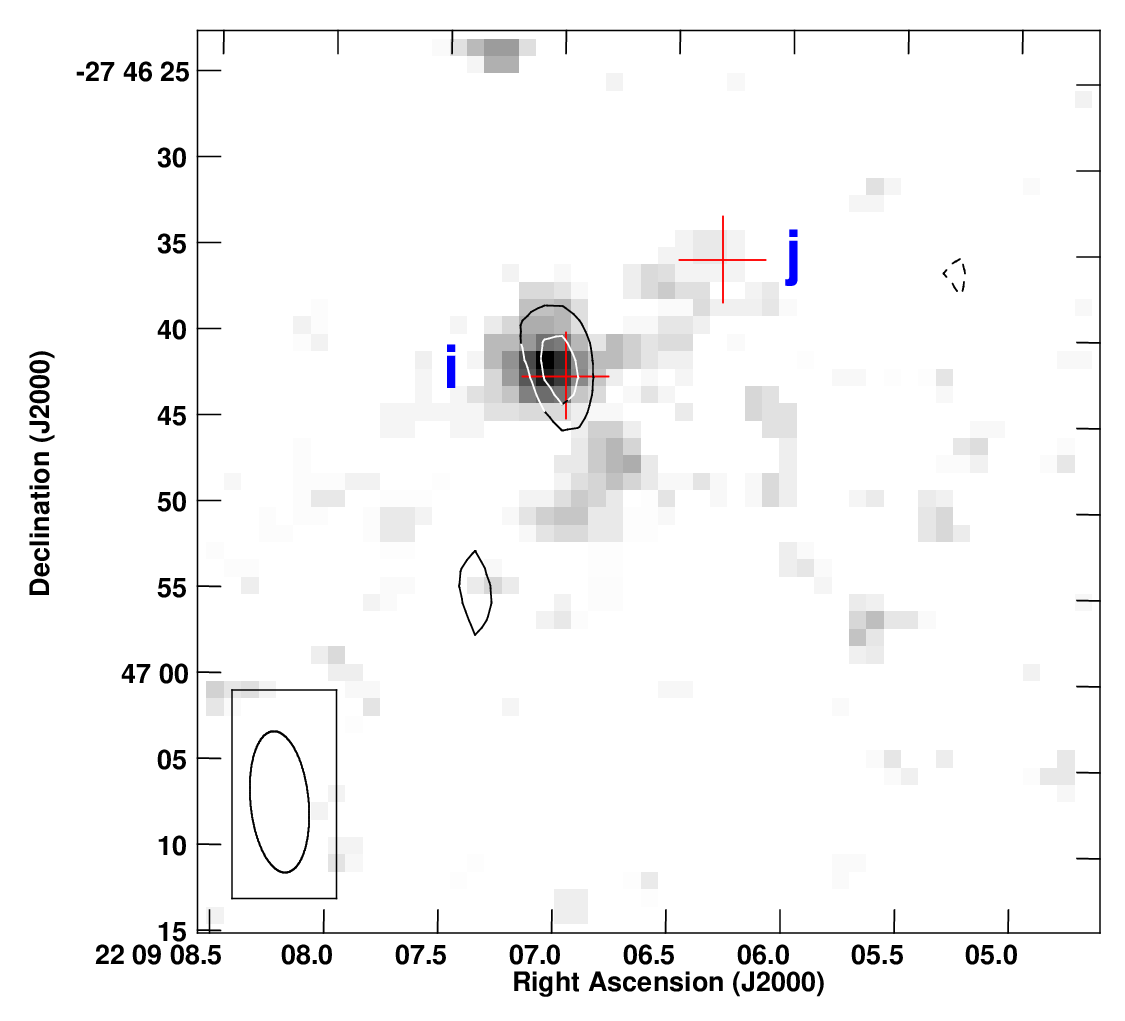}
\caption{\label{fig:h91_int}Maps of the radio emission of HCG\,91 in the vicinity of TDG candidates \textbf{i} and \textbf{j}. Radio contours overlaid on a (high-contrast) cutout of the POSS-II Johnson's red map. The levels are 3, 5, 10 $\times$ r.m.s. noise level. \textbf{Top:} 1.4\,GHz emission; r.m.s. noise level of 250\,$\mu$Jy/beam, angular resolution of 12.8\arcsec$\times$8.1\arcsec. \textbf{Bottom:} 4.86\,GHz emission; r.m.s. noise level of 125\,$\mu$Jy/beam, angular resolution of 8.3\arcsec$\times$3.4\arcsec.}      
\end{figure}

Datasets for the second group, HCG\,91, are noticeably shorter in time (see Table~\ref{tab:data}); this is reflected in the highest noise level among all datasets at both L- (225\,$\mu$Jy/beam) and C-band (125\,$\mu$Jy/beam). HCG\,91 has also the highest number of possible TDGs among those investigated in this study -- 10. Out of this number, 4 are possible radio emitters: HCG\,91c and HCG\,91d are immersed in the same radio structure as their parent galaxies, while HCG\,91i and HCG\,91j share a common radio contour that does not connect to their mother galaxy (Fig.~\ref{fig:h91_main}, upper panel). The higher resolution 1.4\,GHz data (Fig.~\ref{fig:h91_main}, middle panel) show that only HCG\,91i seems to be a (point-source) radio emitter, with a flux density of $2.21 \pm 0.23$\,mJy/beam. This source is still visible at 4.86\,GHz (Fig.~\ref{fig:h91_main}), where its flux density is equal to $0.72 \pm 0.13$\,mJy/beam. Closer inspection (Fig.~\ref{fig:h91_int}) reveals that both at 1.4, and 4.86\,GHz, the radio contours perfectly match with the position of HCG\,91i reported by \citet{eigenthaler15}, suggesting that it is, indeed, a radio-emitting TDG candidate.\\

%=============================
%=============================

%=============================
%=============================

%=============================
%=============================

In case of the last object in this study, HCG\,96, the datasets are both the longest and have the highest resolution among the ones used in this study. The radio contours from the NVSS (Fig.~\ref{fig:h96_main}, upper panel) encompass also the candidate HCG\,91a, while HCG\,91b is not enclosed. Archive data at both 1.4 (Fig.~\ref{fig:h96_main}, middle panel), and 4.86\,GHz (Fig.~\ref{fig:h96_main}, lower panel) show no emission outside of the parent galaxies.

\section{Discussion}
\label{section:discussion}

As most of the relations regarding star formation used in the literature data relevant for this work have been calculated under the assumption of a Salpeter Initial Mass Function (IMF, \citealt{salpeter55}), this IMF was adopted for all the measurements presented in this paper. In particular, the equations provided by \citet{murphy11}, which use the Kroupa IMF \citep{kroupa01}, where re-calibrated by introducing a factor of 1.6, as outlined by \citet{calzetti07}. Common values for the distances to the sources relevant for this study were also adopted, and all values dependant on them were re-calculated.

\subsection{What is the origin of the radio emission?}
\label{section:discussion:origin}

Analysis of the archive data suggests that among the TDG candidates from HCG\,26, HCG\,91, and HCG\,96, only HCG\,91i is visible at 1.4 and 4.86\,GHz. In order to extract the exact coordinates of the radio source, the \textsc{aips} task \textsc{imfit} was used. The derived positions, as well as the position of the $\rm H_{\alpha}$ emitter associated with HCG\,91i given by \citet{eigenthaler15} are presented in Table~\ref{tab:positions}. It turns out that the offset between the fitted maxima of radio emission at 1.4 and 4.86\,GHz is equal to 1.2\arcsec. The theoretical precision (returned by the \textsc{imfit} task) is around 0.6\arcsec, so this displacement is not worrying -- moreover, as the beam size at each of the frequencies is larger, it is in fact negligible. The exact distance between the detection at 4.86\,GHz (where the resolution is higher) and the position estimated from $\rm H_{\alpha}$ data is even lower -- 0.3\arcsec. Assuming that the whole system is located approximately 101\,Mpc away \citep{hickson92}, this translates into a physical separation of less than 150\,pc. HCG\,91j, the other nearby TDG candidate, is located as far as 11.3\arcsec -- a distance larger than the size of the beam at 4.86\,GHz. This altogether suggests that if the radio emission is intrinsic to any of the TDG candidates, then HCG\,91i would be its most probable host.

\begin{table}[htp]
\caption{\label{tab:positions}Positions of the radio and $\rm H_{\alpha}$ sources associated with HCG\,91i.}
\begin{center}
\begin{tabular}{lcc}
\hline
\hline
Data 		    		& $\alpha$ (J2000.0)    & $\delta$ (J2000.0)	            \\
\hline
$\rm H_{\alpha}^{1}$	&	$22^h 09^m 06.98^s$	&	$-27^{\circ} 46' 42.4'' $   	\\
1.4\,GHz			    &	$22^h 09^m 06.95^s$ &	$-27^{\circ} 46' 43.0'' $   	\\
4.8\,GHz			    &	$22^h 09^m 07.00^s$	&	$-27^{\circ} 46' 42.2'' $   	\\
\hline
\end{tabular}
\end{center}
1) From \citet{eigenthaler15}
\end{table}

While HCG\,91i is the single detected radio source among all of the TDG candidates studied, it is not the most luminous one in the $\rm H_{\alpha}$ line \citep{eigenthaler15}. HCG\,91c, HCG\,91d, HCG\,91e, and HCG\,91j have all higher $\rm H_{\alpha}$ luminosity, while HCG\,91a and HCG\,91g are similar to the detected TDGc in this regard. As the $\rm H_{\alpha}$ star formation rate $\rm (SFR_{H_{\alpha}}$) is proportional to the $\rm H_{\alpha}$ luminosity \citep{murphy11}, the most luminous object should also be the most actively star forming one. Moreover, the 1.4\,GHz emission can also be used as an estimator of the SFR \citep{condon02A,murphy11, heesen14}, so it is possible both to estimate the expected 1.4\,GHz luminosity, and the 1.4\,GHz-derived SFR ($\rm SFR_{1.4 GHz}$) -- and compare these values to each other. In order to calculate $\rm SFR_{1.4 GHz}$ and to estimate the expected radio flux at that frequency, Equation (17) from \citet{murphy11}, re-calibrated to the Salpeter IMF \citep{salpeter55}, connecting the radio luminosity and star formation rate was used:

\begin{equation}\label{eq:SFR_radio_murphy}
\bigg(\frac{\rm SFR_{1.4 GHz}}{M_{\odot}\rm yr^{-1}} \bigg) = 1.02 \times 10^{-28} \bigg(\frac{L_{\rm 1.4 GHz}}{\rm erg s^{-1} Hz^{-1}}\bigg)
\end{equation}

Substituting $L_{\rm 1.4 GHz} = 4\pi d^{2} S_{\rm 1.4 GHz}$, where $d$ is the distance to the source, and expressing $S_{\rm 1.4 GHz}$ in [Jy], and $d$ in [Mpc], this equation takes the following form:

\begin{equation}\label{eq:SFR_radio_mine}
\bigg(\frac{\rm SFR_{1.4 GHz}}{M_{\odot}\rm yr^{-1}} \bigg) = 1.22\times10^{-1} \bigg(\frac{S_{\rm 1.4 GHz}\cdot d^{2}}{\rm Jy\,Mpc} \bigg)
\end{equation}

To estimate the radio flux from $\rm SFR_{H_{\alpha}}$,  this can be transformed into:

\begin{equation}\label{eq:flux_radio_mine}
\bigg(\frac{S_{\rm 1.4 GHz}}{\rm Jy}\bigg) = 13.08\bigg( \frac{\mathrm{SFR_{H_{\alpha}}}\cdot d^{-2}}{\rm M_{\odot}\rm yr^{-1}Mpc^{-2}}\bigg)
\end{equation}

Equations \ref{eq:SFR_radio_mine} and \ref{eq:flux_radio_mine} where then used to calculate the expected SFR/upper constraints from the radio flux/3 r.m.s. noise level, and vice versa. Apart from the detected HCG\,91i, the most $\rm H_{\alpha}$-luminous TDG candidates in each of the systems were used. Results can be seen in Table~\ref{tab:SFR}. It turns out, that \emph{none} of the TDG candidates should be detected at 1.4\,GHz! The radio flux extrapolated from the $\rm H_{\alpha}$ data for most of these objects should not exceed a few $\mu$Jy in most of the cases -- thus being undetectable using the currently available radio data. The radio upper limits for the SFR of the non-detected TDGs are approximately two orders of magnitude higher, than that from \citet{eigenthaler15}. The same conclusion is derived when using equation 21 from \citet{condon92}: assuming a spectral index of 0.8, none of the TDG candidates has an expected radio flux higher than approximately 50\,$\mu$Jy. Such a discrepancy raises doubts if the radio emission is indeed connected to the TDG candidate; another, straightforward explanation might be that it originates in a background radio source, which is just by chance so angularly close to the $\rm H_{\alpha}$ maximum associated with HCG\,91i. Therefore, a series of different tests have been carried out, all of them presented in paragraphs below. To avoid excessive usage of terms like "radio emission", "radio source" etc. the designation HCG\,91RS will be used from now on to name the radio counterpart. 

\begin{table}[htp]
\caption{\label{tab:SFR}Radio flux density, its extrapolation from the $\rm H_{\alpha}$ luminosity, and a comparison between the 1.4\,GHz, and $\rm H_{\alpha}$-derived SFR for HCG\,91i and three non-detected TDG candidates}
\begin{center}
\begin{tabular}{lcccc}
\hline
\hline
Object 	&  $\frac{S_{\rm 1.4GHz}}{\rm \mu Jy}$ & $\frac{S_{\rm 1.4GHz, est.}}{\rm \mu Jy}$ & $\frac{\rm log(SFR_{1.4 GHz})}{M_{\odot}\rm yr^{-1}}$ & $\frac{\rm log(SFR_{H_{\alpha}})}{\rm M_{\odot}\rm yr^{-1}}$\\
\hline
HCG 91i	&	2210	&	7.3		&	0.41	&	--2.27	\\
HCG 91d	&	<675	&	15.7	&	<--0.08	&	--1.94	\\
HCG 26a &	<150	&	2.8		&	<--0.47	&	--2.45	\\
HCG 96a	&	<360	&	1.5		&	<--0.08	&	--2.77	\\
\hline
\end{tabular}
\end{center}
\end{table}

\subsection{Radio spectrum of HCG\,91RS}
\label{section:discussion:radio}

A possibility that HCG\,91RS is just an imaging artefact would be one of the most straightforward explanations. However, the detection was made at two separate observing bands; moreover, at the L-band, observations were carried out using two different configurations of the VLA. As a result, the point spread functions of each of the datasets are different, so emergence of an artefact at the same sky position is unlikely. The flux densities of HCG\,91RS derived both from the NVSS and the high resolution data are also coherent. Therefore, this scenario seems not to be a probable one.\\

Additional information on the nature of a radio source can be derived from its radio spectrum. Unfortunately, HCG\,91RS is not an easy target for such a study: it has a relatively steep spectrum, which hinders efforts to detect it at higher frequencies (especially as it is a rather weak emitter). In addition, in order to avoid smearing its signal with the emission from the central galaxy of HCG\,91, high resolution observations are desirable. The NRAO Data Archive lists one possibly useful dataset -- recorded at 8.46\,GHz. Assuming the same spectral index as between L- and C-bands -- $0.91 \pm 0.21$ ($\alpha \propto \nu^{-\alpha}$) -- HCG\,91RS should have a flux density of $0.45 \pm 0.13$\,mJy at 8.46\,GHz -- detectable at at least 5\,r.m.s. level. However, this is not the case; either HCG\,91i is extended at the angular scales larger than that of the beam (which is 0.5\,arcsec in this case) and the resolved parts are too weak to be detected, or its spectrum steepens even further between 4.86 and 8.46\,GHz -- the flattest slope to imply a lack of detection at the given 3\,r.m.s level is 1.41.\\

\begin{figure}[htp]
\centering
    \includegraphics[width=0.4\textwidth]{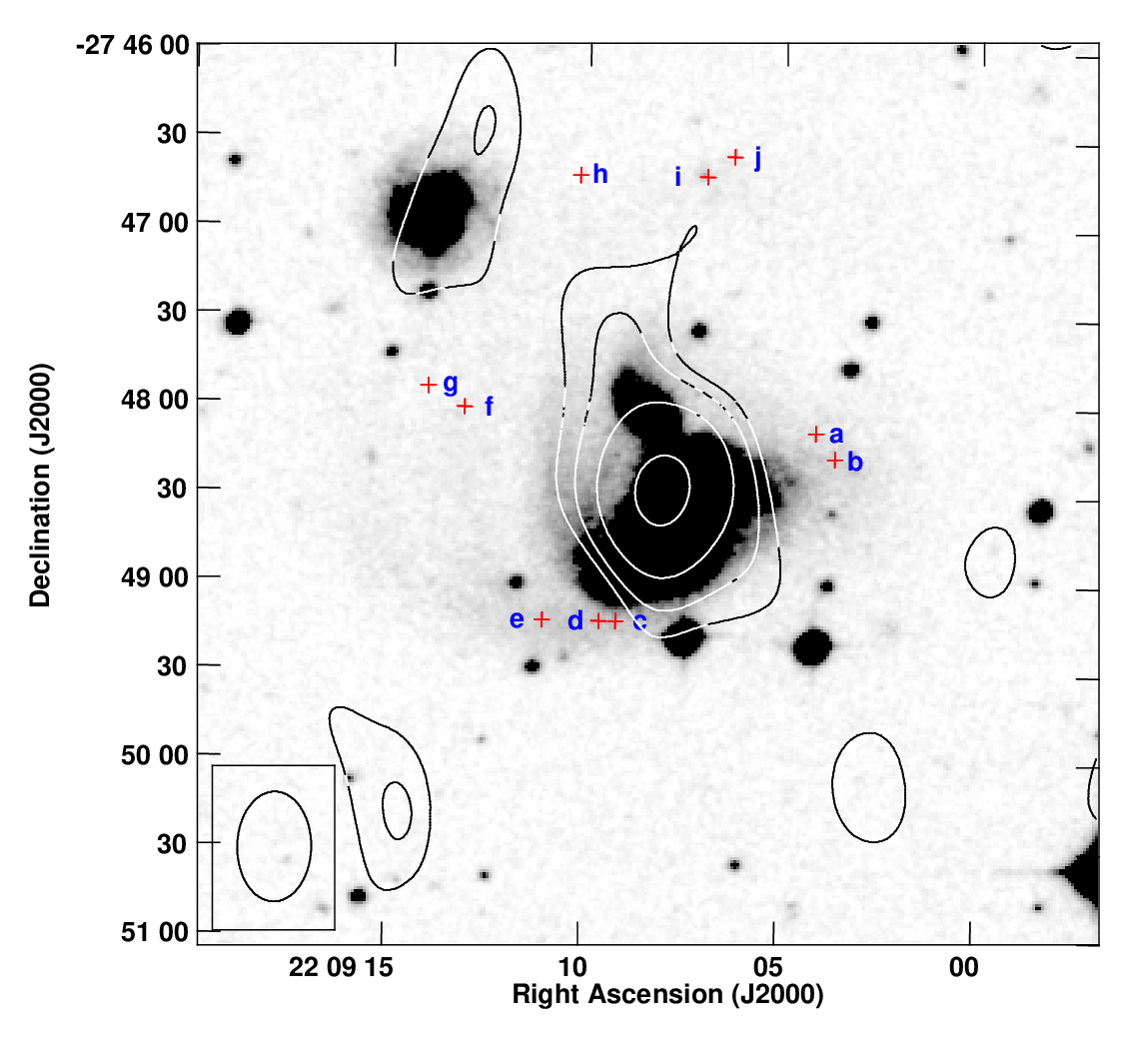}
\caption{\label{fig:h91_tgss}Map of the radio emission of HCG\,91 at 150\,MHz. TGSS radio contours overlaid on a (high-contrast) cutout of the POSS-II Johnson-R filter map. The levels are 3, 5, 10, 25 $\times$ r.m.s. noise level of 1.5\,mJy. The angular resolution is 38.8\arcsec$\times$25\arcsec.}     
\end{figure}

HCG\,91\,i can be looked upon at the lower frequencies, too -- in the TGSS ADR \citep{tgssadr} survey; this data were collected at 150\,MHz, have an angular resolution of 25\arcsec on 25\arcsec$\times \rm sec(-19^{\circ})$, but come with a limited sensitivity to extended structures. Assuming a 1420\,MHz flux density of HCG\,91RS of $2.24 \pm 0.23$\,mJy, and once again, $\alpha = 0.91 \pm 0.21$, one can expect the 150\,MHz flux density to be around 20\,mJy; this is, at a 13$\times$r.m.s. level. However, similarly to the 8.46\,GHz data, this is not the case. The TGSS ADR map (see Fig.~\ref{fig:h91_tgss}) reveals emission from the central pair of galaxies only. If the 3$\times$r.m.s. level is substituted to calculate the spectral index between 150\,MHz and 1.4\,GHz, then the resulting value is equal to $0.31 \pm 0.06$. This would indicate a very flat spectrum, reaching the flatness limit for the non-thermal radio spectra \citep{weiler88}. As a result, either the radio spectrum steepens significantly in the supra-GHz regime, but is relatively flat at lower frequencies, or the low frequency flux density is underestimated.\\

%=============================
%=============================

Gathering the radio information altogether, there are two likely explanations for the spectrum of HCG\,91RS. If it is a background radio galaxy, it is a somewhat older object, with a spectrum quickly steepening around 1\,GHz, or an example of a Gigahertz-Peaked Source \citet{odea91}. If the emission indeed comes from HCG\,91i, then the possible mechanism responsible for the low frequency flattening could be the absorption of the synchrotron radiation on the thermal electrons in the source itself -- a process that is expected to take place in the star-forming regions. With TDG candidates being in fact detached, self-gravitating regions of excessive star formation -- and given the estimated $\rm SFR_{1.4 GHz}$ for HCG\,91RS -- this could explain the lack of detection at 150\,MHz. Similar situation happens for the TDG candidate SQ--B from HCG\,92: assuming the flux density and spectral index provided by \citet{bnw13B}, one could expect it to be detectable in the TGSS, which is not the case. At the higher frequency, the lack of the detection of HCG\,91i can be attributed to the over-resolving of the source's structure: the 0.5-arcsec resolution of the set translates into a physical scale of around 250\,pc -- smaller than the typical size of a TDG candidate.

\subsection{Dust attenuation}
\label{sec_disc_dust}

Given the fact that in case of HCG\,91i and 91RS it is the $\rm SFR_{H_{\alpha}}$ that is much lower than $\rm SFR_{1.4 GHz}$, a possible explanation for the mismatch between these two indicators could be the extinction of $\rm H_{\alpha}$ flux on dust grains. This effect can be significant, and usually can be corrected for by using either the 22, 24 or the 25\,$\mu$m data (see eg. \citealt{calzetti07, murphy11, kennicutt12}). It would be even more plausible reason given the fact that HCG\,91i is located at the tip of a tidal arm/tail -- such TDGs are prone to accumulate larger amounts of matter, including dust \citep{bournaud04}. There is no direct detection of 22\,$\mu$m emission from HCG\,91i in the WISE data, but some upper constraints can still be derived. Assuming an upper flux constraint of $\approx 2.3$\,mJy calculated from the catalogue data on the magnitudes of the sources of interest following the instructions provided in the \textit{Explanatory Supplement to the WISE Preliminary Data Release Products
\footnote{http://wise2.ipac.caltech.edu/docs/release/prelim/expsup/wise\_prelrel\_toc.html}
} and using Equations 6 and 7 from \citet{murphy11} (re-calibrated into the Salpeter IMF), one arrives at $\rm SFR_{H_{\alpha,corr}} = 0.1\,\rm M_{\odot}\rm yr^{-1}$. This value is more than an order of magnitude higher than the non-corrected $\rm SFR_{H_{\alpha}}$from \citet{eigenthaler15}; however, this is the theoretical maximal expected flux, and yet it is still about an order of magnitude lower, than $\rm SFR_{1.4 GHz}$. Therefore, dust extinction of $\rm H_{\alpha}$ flux solely can't explain the observed discrepancy.

\subsection{Typical discrepancies between $\rm SFR_{1.4 GHz}$ and $\rm SFR_{H_{\alpha}}$ for starbursting dwarf galaxies}

The most challenging issue in linking HCG\,91i to HCG\,91RS is the discrepancy between its $\rm SFR_{1.4 GHz}$ and $\rm SFR_{H_{\alpha}}$. To investigate if such a situation can be regarded as a typical one, I have used Eq.\ref{eq:SFR_radio_mine} for a subset of nearby starbursting dwarf galaxies from \citet{mcquinn10} for which the 1.4\,GHz NVSS data \citep{nvss} were available. To supplement this sample, two TDG candidates SQ--A and SQ--B were added. Distances were taken from the NASA Extragalactic Database (except for SQ--A and SQ--B, where \citealt{hickson92} estimates were used). 
All of these objects are small galaxies, most of them have a diameter lower than 3\,kpc (larger sizes of TDG candidates might arose from a less precise measurement). Results can be seen in Table~\ref{tab:SFR_other}. It turns out that in general, those star-bursting dwarf galaxies that have significant 1.4\,GHz flux have $\rm SFR_{1.4 GHz}$ in agreement with $\rm SFR_{H_{\alpha}}$. Several objects have their radio estimates about an order of magnitude lower, than that from $\rm H_{\alpha}$; however, these are all relatively weak objects (in both regimes). In addition, NGC\,625 and NGC\,2366 are the smallest ones considered in this sample. As a result, uncertainties in radio flux can easily lead to a large mismatch between calculated SFRs.\\
In case of SQ--A and SQ--B, the radio data have much higher resolution and sensitivity -- hence, it should be possible to avoid such uncertainties. However, while for SQ--B both SFR estimates are in a nearly perfect agreement, that does not happen in the case of SQ--A. Here the $\rm SFR_{H_{\alpha}}$ is very high; more than two times higher than the estimated $\rm SFR_{1.4 GHz}$.
Even more interestingly, the tendency observed for HCG\,91i and HCG\,91RS is reversed: for SQ--A it is the $\rm SFR_{H_{\alpha}}$ measurement (dust-corrected), that is much higher than the radio one. Unfortunately, the lack of other known radio-emitting TDG candidates renders it impossible to investigate if discrepancies between radio and $\rm H_{\alpha}$ estimators are a common occurrence. Nevertheless, none of the comparison objects shows as large difference between these two estimators as HCG\,91i and HCG\,91RS do.

\begin{table}[htp]
\caption{\label{tab:SFR_other} $\rm SFR_{1.4 GHz}$ for a subset of starburst dwarf galaxies derived from the NVSS catalogue and compared to $\rm SFR_{H_{\alpha}}$ from \citet{mcquinn10}}
\begin{center}
\begin{tabular}{lccccc}
\hline
\hline
Object 	& $ \frac{\rm d}{\rm Mpc}$$^1$ & $ \frac{\rm D}{\rm kpc}$ & $\frac{S_{\rm 1.4GHz}}{\rm mJy}$ & $\frac{\rm SFR_{H_{\alpha}}}{M_{\odot}\rm yr^{-1}}$ & $\frac{\rm SFR_{1.4 GHz}}{\rm M_{\odot}\rm yr^{-1}}$\\
\hline
HCG 91i/RS$^{2}$ & 98  & 3.15 & 2.2 & <0.1 & 2.56 \\
IC  4662         & 4.4 & 1.37 & 40  & 0.08 & 0.09 \\
NGC  625         & 2.8 & 0.87 & 10  & 0.04 & 0.01 \\
NGC  784         & 5.0 & 1.55 & 3   & 0.12 & 0.01 \\
NGC 1569         & 3.4 & 1.06 & 362 & 0.24 & 0.51 \\
NGC 2366         & 1.5 & 0.47 & 20  & 0.16 & 0.01 \\
NGC 4214         & 7.5 & 2.34 & 38  & 0.13 & 0.26 \\
NGC 4449         & 5.8 & 1.81 & 270 & 0.97 & 1.11 \\
NGC 5253         & 9.3 & 2.90 & 86  & 0.40 & 0.91 \\
SQ--A$^{3}$      & 90  & 3.59 & 0.8 & 1.84 & 0.79 \\
SQ--B$^{4}$      & 90  & 3.32 & 0.6 & 0.57 & 0.59 \\
\hline 
\end{tabular}
\end{center}
$^{1}$ Diameter derived from \textit{logd25} parameter from \citet{hyperleda} (if not stated otherwise);\\
$^{2}$ Radio flux taken from this work, $\rm SFR_{H_{\alpha}}$ taken from \citet{eigenthaler15}, recalculated using consistent distance estimate, and corrected for the maximal dust contamination, optical diameter roughly estimated using the background DSS map;\\
$^{3}$ Radio flux taken from \citet{xu03}, $\rm SFR_{H_{\alpha}}$ taken from \citet{xu03} and recalculated using consistent distance estimate, optical diameter roughly estimated using the WISE Band 1 data;\\
$^{4}$ Radio flux taken from \citet{xu03}, $\rm SFR_{H_{\alpha}}$ taken from \citet{lisenfeld16} and recalculated using consistent distance estimate, and re-calibrated from Kroupa to Salpeter IMF, optical diameter roughly estimated using the WISE Band 1 data.
\end{table}

\subsection{Stellar content and SFR--M$_{*}$}
\label{section:discussion:content}

Not only is there a large discrepancy between $\rm SFR_{1.4 GHz}$ and $\rm SFR_{H_{\alpha}}$, but also the very value of radio SFR seems to be unrealistically high: objects listed by \citet{lisenfeld16} exhibit star formation rates of $0.005-0.32 \rm M_{\odot}\rm yr^{-1}$, so at least an order of magnitude lower than $\rm SFR_{1.4 GHz} = 2.6 \rm M_{\odot}\rm yr^{-1}$ for HCG\,91RS. However, there are at least two known examples of TDG candidates with a relatively high SFR, namely SQ--A and SQ--B regions present in HCG\,92. In order to use a consistent distance estimate for all the derivations carried out in this study, I have recalculated the $\rm SFR_{H_{\alpha}}$ provided by \citet{xu03} for SQ--A and \citet{lisenfeld16} for SQ--B, arriving at 1.84 and 0.57 $\rm M_{\odot}\rm yr^{-1}$, respectively. It is then feasible to test if such values would be reliable, eg. by comparing the derived SFR to the stellar content of these objects. In order to do so, I have used the WISE 3.6 and 4.5\,$\mu$m data for all three objects and calculated the stellar content using the following relation from \citet{eskew12}. The infrared fluxes were calculated using the same instructions as in \ref{sec_disc_dust}:
\begin{equation}
\rm \bigg(\frac{M_{*}}{M_{\odot}}\bigg)=10^{5.65}\bigg(\frac{(F_{3.6\mu m})^{2.85}}{Jy}\bigg)\bigg(\frac{(F_{4.5 \mu m})^{-1.85}}{Jy}\bigg)\bigg(\frac{20\times D}{Mpc}\bigg)^2
\end{equation}{}

This yields stellar masses equal to $10^{9.04} \rm M_{\odot}$ for HCG\,91i, $10^{9.37} \rm M_{\odot}$ for SQ--A, and $10^{8.58} \rm M_{\odot}$ for SQ--B. This in turn identifies SQ--A and HCG\,91i as relatively massive TDG candidates, as the statistical study of \citet{kaviraj12} shows that an average stellar mass of such an object varies between $10^{8.2}$ and $10^{8.4} \rm M_{\odot}$, depending on its location (at the base/along/ at the tip of the tidal tail). These values were then substituted to the SFR--mass relation for late type galaxies given by \citet{calvi18}. The range of expected SFR (in $\rm M_{\odot}\rm yr^{-1}$) is 0.2--1.3 for HCG\,91i, 0.3--1.6 for SQ--A, and 0.1--0.6 for SQ--B. In case of SQ--A and SQ--B, it is in a good agreement with both $\rm SFR_{1.4 GHz}$ and $\rm SFR_{H_{\alpha}}$; the $\rm SFR_{H_{\alpha}}$ estimate for the former one is a bit larger than the expected value. Contrary to that, for HCG\,91i, only one SFR indicator can be assumed to be similar to the predictions based on the the stellar mass -- the radio one, albeit the derived SFR is nearly twice as much as the expected one. On the other hand, $\rm SFR_{H_{\alpha}}$ for HCG\,91i, even after correcting for the dust contamination, is more than an order of magnitude lower, than what one would expect from the SFR--M$_{*}$ relation. As a result, not only is the star formation rate as high as the estimated $\rm SFR_{1.4 GHz}$ feasible, but it turns out to be the one fitting better to the SFR-M$_{*}$ relation.

\subsection{Variations in the SFR over the time}

Another explanation of the discrepancy between $\rm SFR_{1.4 GHz}$ and $\rm SFR_{H_{\alpha}}$ could be a series of star formation bursts inside HCG\,91i. SFR estimators are sensitive to stars of different age: as outlined by \citet{kennicutt12}, for the $\rm H_{\alpha}$ estimator the mean age of the stars that contribute to the emission is around 3\,Myrs, and 90\% of emission comes from the ones that are younger than 10\,Myrs. Other estimators are sensitive to different populations: for example, the ultraviolet tracer is sensitive to the population which has a mean age of 10\,Myrs, and the age range of the objects that contribute to this emission is much larger, than in the case of $\rm H_{\alpha}$: it is 100\,Myrs for the far- and 200\,Myrs for the near-ultraviolet light. In case of the 1.4\,GHz emission, the mean age is equal to 100\,Myrs -- and the "boundary" age is even not estimated. Series of subsequent bursts could result in an enhanced magnetic field/radio emission, suggesting a still high SFR, while the most current value, traced by the $\rm H_{\alpha}$ emission would be lower, if the object under consideration is in a more quiescent phase. If this is indeed the case of HCG\,91i, then it should still be a strong UV emitter. In order to evaluate this scenario, I have checked the GALEX FUV and NUV maps. The results are not encouraging: first of all, there is no UV source that could be directly associated with HCG\,91i. The closest ones are very faint (NUV fluxes of less than 15\,$\mu$Jy), and only one of them was detected in both FUV and NUV. The distance between them and the radio maximum is significant -- more than 10 arcseconds, so at least 4.5\,kpc -- more than the expected size of the whole TDG candidate. I have estimated the SFR using the relations presented by \citet{kennicutt12}, derived from the earlier works od \citet{hao11} and \citet{murphy11}, and recalibrated for the Salpeter IMF. The FUV and NUV fluxes were taken for the nearest source detected in both of these bands, and corrected for the dust extinction using the WISE Band 4 upper flux limit of 2.3\,mJy. The dust-enshrouded $\rm SFR_{FUV}$ is equal to approximately $\rm 0.03 M_{\odot}\rm yr^{-1}$, and $\rm SFR_{NUV}$ is approximately $\rm 0.04 M_{\odot}\rm yr^{-1}$. These values are lower than those estimated for the dust-enshrouded $\rm SFR_{H_{\alpha}}$ (and thus, much lower than the $\rm SFR_{1.4 GHz}$ ). Therefore, the hypothesis of subsequent burst can be safely discarded.

\subsection{An inherited magnetic field?}

One of the most fundamental differences between the $\rm SFR_{1.4 GHz}$ and $\rm SFR_{H_{\alpha}}$ estimators is the evolutionary stage of the objects they probe. The latter one relies on the presence of an ionised gas surrounding a young, massive star. Contrary to that, the radio one is aimed at a much older ones: it is, in fact, a \textit{post mortem} survey, as the synchrotron emission relies on the relativistic electrons supplied by the supernovae. Traces of these processes fade away on different timescales: radio emission can still be detectable tens of Myrs after last electrons where supplied. Therefore, one can propose a hypothesis that while both the radio and $\rm H_{\alpha}$ emission are connected to HCG\,91i, only the latter one is caused by processes intrinsic to (or triggered inside) this object. Majority of the radio emission is caused by an electron population supplied to the magnetic field when what is now regarded as a TDG candidate was still a part of its parent object. Analysis of the radio spectrum of HCG\,91RS seems to be consistent with this scenario. With a spectral index of $0.91 \pm 0.21$, it is a rather steep (thus aged) one. Young supernovae remnants have in general much flatter indices: the lower limit is assumed to be equal to 0.3 \citep{weiler88}. Such flat spectra are indeed seen in case of disk star forming regions -- like it happens in case of the dwarf starburst galaxy pair Arp\,269 \citep{bnw16}. A rough estimate of the spectral age, made under the assumption that the break frequency of the spectrum is already below 4.86\,GHz suggests that the radio emission must be older than approximately 10-15\,Myrs -- likely more. 
In addition, the aforementioned scenario easily explains the excessive strength of the magnetic field associated with HCG\,91RS (see Sect.~\ref{section:discussion:mfield}). The obtained value of 11--16 $\mu$G is, as already mentioned, similar to that of example disk star forming regions. If HCG\,91i began its life as a region of vigorous star formation inside the spiral arm of HCG\,91A and was later separated due to the action of tidal forces, then a strong, "remnant" magnetic field would be indeed expected.
While at the first glance the scenario where HCG\,91i and HCG\,91RS are the same object, but their emission represents fundamentally different moments in its past seems to be able to easily explain the observed discrepancies, it also has several important caveats. There is neither information on the interaction history of the whole group, nor about the stellar population of HCG\,91i. It is then impossible to address the problem quantitatively, eg. compare the spectral age with the expected time elapsed since the interaction started. Additional studies on HCG\,91i and the history of the system as a whole are necessary to test the feasibility of this hypothesis.

An additional hint for the feasibility of the hypothesis described above comes from the analysis of optical maps: albeit there are four more $\rm H_{\alpha}$--luminous TDG candidates in their host system than HCG\,91i, it is the only one luminous enough to be included in the USNO A-2.0 catalogue with an absolute magnitude in these filters of $\approx -16$, and the only one to be unambiguously detected in the infrared 3.6 and 4.5\,$\mu$m data. Even in the background maps used for the radio images, it is the single TDG candidate with a clear optical counterpart. However, again there is a possibility that the observed object is just a superposition of a TDG candidate, and a distant background source.

Alas, the identification of HCG\,91RS with HCG\,91i remains ambiguous. Whereas there is neither a certain explanation for the discrepancy between $\rm SFR_{1.4 GHz}$ and $\rm SFR_{H_{\alpha}}$, nor the possibility of a background AGN galaxy can be ruled out fully, analysis of the stellar content and its relationship with the system's expected SFR clearly suggests that values similar to $\rm SFR_{1.4 GHz}$ would be expected. Additional data would be needed to prove, or disprove the hypothesis that HCG\,91i is indeed another example of a rare, starbursting tidal dwarf galaxy candidate possessing a detectable magnetic field.

Last but not least, it is also worth to mention that many of the arguments presented in this study could be better assessed if there was a considerable literature data on radio emitting TDG candidates. With only three of such objects being known (assuming that HCG91\,i is one of these), it is impossible to evaluate if any studied parameter can be considered typical (or not) for this class. A larger study aimed at revealing new radio-emitting TDG candidates is thus desirable.

\subsection{Magnetic field inside the TDG candidates}
\label{section:discussion:mfield}

If HCG\,91RS is indeed the radio counterpart of HCG\,91i, then it is possible to use the flux densities at 1.4 and 4.86\,GHz to estimate the strength of its magnetic field. The spectral index of $0.91 \pm 0.21$ was substituted into the \textsc{bfeld} code \citep{bfeld} that calculates the basic properties of the magnetic field, together with a pathlength of 2000-4000\,pc (similar to the size of the optical counterpart), and the proton-to-electron ratio of 100, which is believed to be maintained even in the starburst galaxies \citep{lacki13}. The estimated magnetic field strength (under the assumption of equipartition of energy between the magnetic field and the cosmic rays) is then 11--16 $\mu$G. This is a strong magnetic field -- stronger than the typical ones found for spiral galaxies by \citet{niklas95}, and approximately two times stronger than the one derived for the radio-emitting TDGc  SQ--A and SQ-B \citep{bnw13B}. Such a strength is similar to that of the compact disk areas of star formation: \citet{beck13} lists several objects that host even stronger magnetic fields, and the study of the magnetic field of starburst dwarf spiral galaxy NGC\,4490 \citep{bnw16} reveals a handful of disk starburst regions hosting magnetic fields exceeding 20\,$\mu$G in strength. Compared to the other dwarf galaxies, HCG\,91i seems to posses a magnetic field similar to that of the starburst dwarf galaxies studied by \citet{chyzy11}.\\

While no detection was made in case of any other TDG candidate from the list of \citet{eigenthaler15}, it is possible to estimate the upper constraints for the strength of magnetic field in these entities. Using the same pathlength as before, and calculating the spectral index using with the 3\,r.m.s. levels of the respective radio data, it turns out that the magnetic fields of 6--8$\mu$G strength can still remain undetected, due to high r.m.s noise levels. Such a value would also apply to HCG\,91i, if the radio emission comes from a background source. A more sensitive study would allow to derive more strict constraints.

\section{Conclusions}
\label{section:conclusions}

This work attempted to detect the radio counterparts of the Tidal Dwarf Galaxy (TDG) candidates in compact galaxy groups HCG\,26, 91, and 96. On the basis of the gathered and analysed material, the conclusions are as follows:

   \begin{enumerate}
      \item There are clear signs of radio emission spatially coincident with the TDG candidate HCG\,91i (emission maxima matched with less than 150\,pc separation) detected in $\rm H_{\alpha} $ line by \citet{eigenthaler15} both at 1.4, and 4.86\,GHz. The detected emission has a steep spectrum, characterised by an index of 0.91$\pm$0.21;
      \item Analysis of the high resolution 8.46\,GHz and TGSS-ADR 150\,MHz radio data yields no radio detection, suggesting that either the radio source is a Gigahertz Peaked Source (or an aged AGN), or a star-forming region, weak at lower frequencies due to the absorption of synchrotron radiation on the thermal electrons, and over-resolved and thus too faint to be seen at 8.46\,GHz;
      \item Even after the $\rm H_{\alpha}$ luminosity is corrected for the dust attenuation (using upper constraints as no detection was made at 22\,$\mu$m), $\rm SFR_{H_{\alpha,corr}}$ is about an order of magnitude lower than $\rm SFR_{1.4 GHz}$;
      \item Comparison of $\rm SFR_{1.4 GHz}$ and $\rm SFR_{H_{\alpha}}$ for a subset of starbursting dwarf galaxies shows that in most of the cases these two values are either concordant, or the difference can be attributed to the accuracy of the measurement, with the single exception of SQ--A tidal dwarf galaxy candidate, where $\rm SFR_{H_{\alpha}}$ is significantly higher, than its $\rm SFR_{1.4 GHz}$ (and HCG\,91i, if the radio emission is intrinsic to it);
      \item Analysis of the stellar content and SFR versus stellar mass relation suggests that both HCG\,91i and SQ--A are massive objects, and their expected SFR values are of the order of $1 \rm M_{\odot}\rm yr^{-1}$; hence, in case of the former object, it is the radio estimate that seems to be more feasible;
      \item Ultraviolet star formation rates, even when corrected for the dust attenuation, are very low, thus ruling out the possibility that the discrepancy between different SFR indicators is due to a burst of star formation;
      \item At the moment the scenario in which HCG\,91i is a former star-forming region of its parent galaxy, with the radio emission being a remnant of this evolutionary stage, seems to fit the observations (and estimations) the best;
      \item Should the radio emission originate in HCG\,91i, then the derived strength of the magnetic field inside is 11 --16 $\mu$G, similar to that found in starburst dwarf galaxies, or disk regions of vigorous star formation;\\
      \item If this is not the case, then the upper constraints for the strength of the magnetic field inside all of the TDG candidates in the studied systems vary from 6--8$\mu$G, leaving a possibility that strong magnetic fields can still be left undetected.
   \end{enumerate}

\begin{acknowledgements}
I would like to thank the anonymous referee for a number of comments and suggestions that helped to significantly improve this paper. I am also indebted to Krzysztof Chy\.zy, Marek Jamrozy and Marian Soida from the Astronomical Observatory of the Jagiellonian University for useful comments and suggestions that also helped to improve this paper.
\end{acknowledgements}

%-------------------------------------------------------------------

\allauthors


\begin{thebibliography}{}

% SDSS - DR6
\bibitem[\protect\citeauthoryear{Abazajian et al.}{2009}]{sdss6} 
Adelman-McCarthy, J.~K., et al., 2007, ApJS, 175, 297

% Magnetic fields - revised equipartition formula
\bibitem[\protect\citeauthoryear{Beck \& Krause}{2005}]{bfeld} 
Beck R., Krause M., 2005, AN, 326, 414

% Magnetic fields - Rainer's review
\bibitem[\protect\citeauthoryear{Beck \& Wielebinski}{2013}]{beck13} 
Beck, R., Wielebinski, R., in: Stars and Stellar Systems, Vol. 5: Galactic Structure and Stellar Populations, ed. G. Gilmore, Springer, Berlin 2013

%TDGs - massive ones
\bibitem[\protect\citeauthoryear{Bournaud}{2004}]{bournaud04} 
Bournaud, F., Duc, P.~-A., Amram, P., Combes, F., Gach, J.~-L., 2004, 425, 813

%TDGs - Hubble time survival
\bibitem[\protect\citeauthoryear{Bournaud}{2010}]{bournaud10} 
Bournaud, F., 2010, Advances Astron., 2010

% TDG - 2004 symposium
\bibitem[\protect\citeauthoryear{Brinks, Duc, \& Walter}{Brinks et al.}{2004}]{brinks04} 
Brinks, E., Duc, P.-A., Walter, F. 2004, Recycling Intergalactic and Interstellar Matter (IAU Symp. 217), ed. P.-A. Duc, J. Braine, \& E. Brinks (San Francisco, CA: ASP), 532

% SFR - SFR for different type of galaxies and SFR-mass relation
\bibitem[\protect\citeauthoryear{Calvi et al.}{2018}]{calvi18} 
Calvi, R., Vulcani, B., Poggianti, B.~M., Moretti, M., Fritz, J., Fasano, G., 2018, MNRAS, 481, 3456

% Star formation rate - estimation
\bibitem[\protect\citeauthoryear{Calzetti et al.}{2007}]{calzetti07} 
Calzetti, D., et al., 2007, ApJ, 666, 870

% NGC 4449 - magnetic field
\bibitem[\protect\citeauthoryear{Chy\.zy et al.}{2000}]{chyzy00} 
Chy{\.z}y K.~T., Beck R., Kohle S., Klein U., Urbanik M., 2000, A\&A, 355, 128 

% Magnetic fields -- dwarf galaxies
\bibitem[\protect\citeauthoryear{Chy\.zy et al.}{2011}]{chyzy11} 
Chy\.zy, K.~T., We\.zgowiec, M., Beck, R., Bomans, D.~J., 2011, A\&A, 529, 94

% Radio emission - star formation
\bibitem[\protect\citeauthoryear{Condon et al.}{1992}]{condon92}
Condon, J.~J., 1992, ARA\&A, 30, 575

% NVSS
\bibitem[\protect\citeauthoryear{Condon et al.}{1998}]{nvss} 
Condon J.~J., Cotton W.~D., Greisen E.~W., Yin Q.~F., Perley R.~A., Taylor G.~B., Broderick J.~J., 1998, ApJ, 115, 1693

% Radio emission - star formation
\bibitem[\protect\citeauthoryear{Condon et al.}{2002}]{condon02A}
Condon, J.~J., Cotton, W.~D., Broderick, J.~J., 2002, AJ, 124, 675

% GPS sources - original paper
\bibitem[\protect\citeauthoryear{O'Dea, Baum \& Stanghellini}{O'Dea et al.}{1991}]{odea91} 
O'Dea, C.~P., Baum, S.~A., Stanghellini, C., 1991, ApJ, 380, 66O

% VCC2062 - original paper
\bibitem[\protect\citeauthoryear{Duc et al.}{2007}]{duc07}
Duc, P.~-A., Braine, J., Lisenfeld, U., Brinks, E., Boquien, M. 2007, A\&A, 475, 187

%HCG - TDGs
\bibitem[\protect\citeauthoryear{Eigenthaler et al.}{2015}]{eigenthaler15} 
Eigenthaler, P., Ploeckinger, S.,  Verdugo, M., Ziegler, B., 2015, MNRAS, 451, 2793

% SFR vs Stellar Mass
\bibitem[\protect\citeauthoryear{Eskew, Zaritsky \& Meidt}{Eskew et al.}{2012}]{eskew12} 
Eskew, M., Zaritsky, D., Meidt, S., 2012, AJ, 143, 139

% Star formation - estimators
\bibitem[\protect\citeauthoryear{Hao et al.}{2011}]{hao11}
Hao, C.-N., Kennicutt, R.~C., Johnson, B.~D., Calzetti, D., Dale, D.~A., Moustakas, J., 2011, ApJ, 741, 124 

% WSRT-SINGS - star formation rates, magnetic fields
\bibitem[\protect\citeauthoryear{Heesen et al.}{2014}]{heesen14}
Heesen, V., Brinks, E., Leroy, A.~K., Heald, G., Braun, R., Bigiel, F., Beck, R., 2014, AJ, 147, 103

% HCG - original catalogue
\bibitem[\protect\citeauthoryear{Hickson}{1982}]{hickson82}
Hickson, P., 1982, \apj, 255, 382

% HCG - dynamical properties, distances
\bibitem[\protect\citeauthoryear{Hickson et al.}{1992}]{hickson92}
Hickson, P., Mendes de Oliveira, C., Huchra, J.~P., et al.\ 1992, \apj, 399, 353

% HCG - TDGs in HCGs
\bibitem[\protect\citeauthoryear{Hunsberger \& Zaritsky}{1996}]{hunsberger96} 
Hunsberger, S.~D., Zaritsky, D., 1996, ApJ, 462, 50

% TGSS ADR - original paper
\bibitem[\protect\citeauthoryear{Intema et al.}{2016}]{tgssadr}
Intema, H.~T., Jagannathan, P., Mooley, K.~P., Frail, D.~A., 2016, A\&A, 598, 78

% TDG: statistical study
\bibitem[\protect\citeauthoryear{Kaviraj et al.}{2012}]{kaviraj12} 
Kaviraj, S., Darg, D., Lintott, C., Schawinski, K., Silk, J., 2012, MNRAS, 419, 70

% Star formation - estimators
\bibitem[\protect\citeauthoryear{Kennicutt \& Evans}{2012}]{kennicutt12} 
Kennicutt, R.~C., Evans, N.~J., 2012, ARA\&A, 50, 531

%NGC 1569 - radio emission, magnetic field
\bibitem[\protect\citeauthoryear{Kepley et al.}{2010}]{kepley10} 
Kepley, A.~A, M\"uhle, S., Everett, J., Zweibel, E.~G., Wilcots, E.~M., Klein, U., 2010, ApJ, 712, 536

%Star formation - Kroupa IMF
\bibitem[\protect\citeauthoryear{Kroupa}{2001}]{kroupa01} 
Kroupa, P. 2001, MNRAS, 322, 231

% Magnetic field - equipartition in starbursting galaxies
\bibitem[\protect\citeauthoryear{Lacki \& Beck}{2013}]{lacki13}
Lacki, B.~C., Beck, R., 2013, MNRAS, 430, 317

% TDG - star formation properties
\bibitem[\protect\citeauthoryear{Lisenfeld et al.}{2016}]{lisenfeld16} 
Lisenfeld, U., Braine, J., Duc, P. A., Boquien, M., Brinks, E., Bournaud, F., Lelli, F., Charmandaris, V., 2016, A\&A, 590, 92

% Hyperleda - 2014 version
\bibitem[\protect\citeauthoryear{Makarov et al.}{2014}]{hyperleda}  
Makarov, D., Prugniel, P., Terekhova, N., Courtois, H., Vauglin, I., 2014, A\&A, 570, 13

%TDGs in M81/M82
\bibitem[\protect\citeauthoryear{Makarova et al.}{2002}]{makarova02}
Makarova, L.~N. et al., 2002, A\&A, 396, 473

% TDG in the Antennae
\bibitem[\protect\citeauthoryear{Mirabel, Dottori \& Lutz}{Mirabel et al.}{1992}]{mirabel92} 	
Mirabel, I.~F., Dottori, H., Lutz, D., 1992,A\&A, 256, 19

%SFR in starburst dwarf galaxies
\bibitem[\protect\citeauthoryear{McQuinn et al.}{2010}]{mcquinn10}
McQuinn, K. et al., 2010, ApJ, 721, 297 

% Star formation - estimators
\bibitem[\protect\citeauthoryear{Murphy et al.}{2011}]{murphy11} 
Murphy, E.~J., et al., 2011, ApJ, 737, 67

% Leo Triplet - Effelsberg polarimetry
\bibitem[\protect\citeauthoryear{Nikiel-Wroczy\'nski et al.}{2013a}]{bnw13A}
Nikiel-Wroczy\'nski, B., Soida, M., Urbanik, M., We\.zgowiec, M., Beck, R., Bomans, D. J., Adebahr, B., 2013, A\&A, 553, 4

% Stephan's Quintet - VLA polarimetry
\bibitem[\protect\citeauthoryear{Nikiel-Wroczy\'nski et al.}{2013b}]{bnw13B} 
Nikiel-Wroczy\'nski, B., Soida M., Urbanik, M., Beck, R., Bomans, D.~J., 2013, MNRAS, 435, 149

% Leo TDG - original work
\bibitem[\protect\citeauthoryear{Nikiel-Wroczy\'nski et al.}{2014}]{bnw14B} 
Nikiel-Wroczy\'nski, B., Soida, M., Bomans, D. J., Urbanik, M., 2014, ApJ, 786, 144

% NGC 4490/85 - GMRT+VLA+Effelsberg total power
\bibitem[\protect\citeauthoryear{Nikiel-Wroczy\'nski et al.}{2016}]{bnw16} 
Nikiel-Wroczy\'nski, B., Jamrozy, M., Soida, M., Urbanik, M., Knapik, J., 2016, MNRAS, 459, 683

% Magnetic fields - average strengths in spiral galaxies
\bibitem[\protect\citeauthoryear{Niklas}{1995}]{niklas95} 
Niklas S., 1995, PhD thesis, University of Bonn

%Star formation - Salpeter IMF
\bibitem[\protect\citeauthoryear{Salpeter}{1955}]{salpeter55} 
Salpeter, E.E. 1955, ApJ, 121, 161

% Stephan's Quintet - discovery
\bibitem[\protect\citeauthoryear{Stephan}{1877}]{stephan77} 
St\'ephan, \`E.~J.-M., 1877, MNRAS, 37, 334

% Radio emission - SNR limit for initial spectrum
\bibitem[\protect\citeauthoryear{Weiler \& Sramek}{1988}]{weiler88} 
Weiler, K.~W., Sramek, R.~A., 1988, ARA\&A, 26, 295

% NGC 3627 - galaxy-dwarf collision
\bibitem[\protect\citeauthoryear{We\.zgowiec, Soida \& Bomans}{We\.zgowiec et al.}{2012}]{wezgowiec12} 	
We\.zgowiec, M., Soida, M., Bomans, D. J., 2012, A\&A, 544, 113

% Stephan's Quintet - VLA high-res
\bibitem[\protect\citeauthoryear{Xu et al.}{2003}]{xu03} 
Xu C.~K., Lu N., Condon J.~J., Dopita M., Tuffs R.~J., 2003, ApJ, 595, 665

% TDG - original work
\bibitem[\protect\citeauthoryear{Zwicky}{1956}]{zwicky56} 
Zwicky, F., 1956, Ergebnisse der Exakten Naturwissenschaften 29, 344


%Adelman-McCarthy, J.~K., Agüeros, M.~A., Allam, S.~S., et al. 2007, ApJS, 175, 297

%Beck, R., Wielebinski, R., 2013 in Planets, Stars and Stellar Systems Vol. 5, by Oswalt, Terry D.; Gilmore, Gerard, ISBN 978-94-007-5611-3. Springer Science+Business Media Dordrecht, 2013, p. 641, https://ui.adsabs.harvard.edu/abs/2013pss5.book..641B/abstract

%Bournaud, F., 2010 in Galaxy Wars: Stellar Populations and Star Formation in Interacting Galaxies ASP Conference Series Vol. 423, proceedings of a conference held 19-22 July 2009 at East Tennessee State University, Johnson City, Tennessee, USA. Edited by Beverly Smith, Nate Bastian, Sarah J. U. Higdon, and James L. Higdon. San Francisco: Astronomical Society of the Pacific, 2010., p.177, https://ui.adsabs.harvard.edu/abs/2010ASPC..423..177B/abstract

%Bournaud, F., Duc, P. -A., Amram, P., Combes, F., Gach, J. -L., 2004, A&A, 425, 813, https://ui.adsabs.harvard.edu/abs/2004A%26A...425..813B/abstract

%Brinks, E., Duc, P.~-A., Walter, F., 2004, in International Astronomical Union Symposium no. 217, held 14-17 July, 2003 in Sydney, Australia. Edited by P.-A. Duc, J. Braine, and E. Brinks. San Francisco: Astronomical Society of the Pacific, 2004., p.532, https://ui.adsabs.harvard.edu/abs/2004IAUS..217..532B/abstract

%Calzetti, D., Kennicutt, R.~C., Engelbracht, C.~W., et al. 2007, ApJ, 666, 870

%Makarova, L.~N., Grebel, E.~K., Karachentsev, I.~D., et al. 2002, A&A, 396, 473

%McQuinn, K.~B.~W., Skillman, E.~D., Cannon, J.~M., et al. 2010, ApJ, 721, 297

%Murphy, E.~J., Condon, J.~J., Schinnerer, E., et al. 2011, ApJ, 737, 67

%Niklas, S., 1995, PhD Thesis, Univ. Bonn, https://ui.adsabs.harvard.edu/abs/1995PhDT........57N/abstract


\end{thebibliography}
\end{document}